\documentclass[
reprint,
superscriptaddress,amsmath,amssymb,
prb,aps,
]{revtex4-1}

\usepackage{graphicx}
\DeclareGraphicsExtensions{.pdf,.eps,.png,.jpg}
\usepackage{dcolumn}
\usepackage{float}
\usepackage{bm}
\usepackage{xspace}
\usepackage{color}
\newcommand{\irte}{IrTe$_2$\xspace}
\newcommand{\RN}[1]{\uppercase\expandafter{\romannumeral#1}}

\begin{document}

\title{Local mechanism of valence bond formation in \irte}

\author{T. Ritschel}
\affiliation{Institute of Solid State and Materials Physics,  TU Dresden, 01069 Dresden, Germany}

\author{Q. Stahl}
\affiliation{Institute of Solid State and Materials Physics,  TU Dresden, 01069 Dresden, Germany}

\author{M. Kusch}
\affiliation{Institute of Solid State and Materials Physics,  TU Dresden, 01069 Dresden, Germany}

\author{J. Trinckauf}
\affiliation{Leibniz Institute for Solid State and Materials Research IFW Dresden, Helmholtzstr. 20, D-01069 Dresden, Germany}

\author{G. Garbarino}
\affiliation{European Synchrotron Radiation Facility, BP 220, F-38043 Grenoble Cedex, France}

\author{V. Svitlyk}
\affiliation{European Synchrotron Radiation Facility, BP 220, F-38043 Grenoble Cedex, France}

\author{M. Mezouar}
\affiliation{European Synchrotron Radiation Facility, BP 220, F-38043 Grenoble Cedex, France}

\author{J. Yang}
\affiliation{Department of Physics, New Jersey Institute of Technology, Newark, NJ 01702, USA}

\author{S.W. Cheong}
\affiliation{Rutgers Center for Emergent Materials and Department of Physics and Astronomy, Rutgers, The State University of New Jersey, 136 Frelinghuysen Road, Piscataway, New Jersey 08854-8019 USA}

\author{J. Geck}
\email[]{jochen.geck@tu-dresden.de}
\affiliation{Institute of Solid State and Materials Physics,  TU Dresden, 01069 Dresden, Germany}
\affiliation{W\"urzburg-Dresden Cluster of Excellence ct.qmat, Technische Universit\"at Dresden, 01062 Dresden, Germany}


\date{\today}

\begin{abstract}
Doped \irte is considered a platform for topological superconductivity and therefore receives currently a lot of interest. In addition, the superconductivity in these materials exists in close vicinity of electronic valence bond crystals, which we explore here by means of high-pressure single crystal x-ray diffraction in combination with density functional theory. 
Our crystallographic refinements provide unprecedented information about the structural evolution as a function of applied pressure up to 42\,GPa. Using this structural information for density functional theory calculations, we show that the valence bond formation in \irte is driven by changes in the Ir-Te-Ir bond angle.  When a valence bond is formed, this bond angle decreases drastically, leading to a stabilization of local valence bonds large enough to push them out of a broad band continuum. This unusual local mechanism of valence bond formation in an itinerant material provides a natural explanation for the different valence bond orders in \irte, implies a very strong electron-phonon coupling and is most likely relevant for the superconductivity as well.
\end{abstract}

\maketitle


\section{Introduction}\label{sec:intro}

Novel quantum states and their mutual interactions are a major topic of contemporary condensed matter science. Famous and intensively studied examples are unconventional superconductivity\cite{Stewart:2017x}, charge density waves\cite{Arpaia:2019aa}, quantum spin liquids\cite{Savary:2017a}, topological states of matter\cite{Yan:2017a} or Dirac materials\cite{Wehling:2014m}. 
Here two branches of research are attracting particularly much attention: The first one is the role of fluctuating electronic order and quantum magnetism for superconductivity. 
The second one, is the combination of topological electrons and superconductivity in the very sough-after topological superconductors.

In this regard, transition metal dichalcogenides (TMDs) of the type $TX_2$ ($T$: transition metal, $X$: chalcogenide) are extremely interesting, as they are in fact notorious for harbouring a variety of intriguing quantum phenomena. To name just a few cases in point, both superconductiviy and charge density waves occur in NbSe$_2$, TaSe$_2$ and TaS$_2$, while WTe$_2$ presently attracts a lot of interest in the context of topological Weyl-physics\cite{Rossnagel:2011a, Yan:2017a, Yang:2017e,Gye:2019s}. 
Typical TMDs can be regarded as stacks of two-dimensional planes with relatively weak interactions between the planes. As a result, the electronic structure  of TMDs is usually very anisotropic and possess a pronounced two-dimensional character, which, in turn, promotes electronic instabilities and unconventional electronic ground states. 
%

\begin{figure*}
    \centering
    \includegraphics[width=\textwidth]{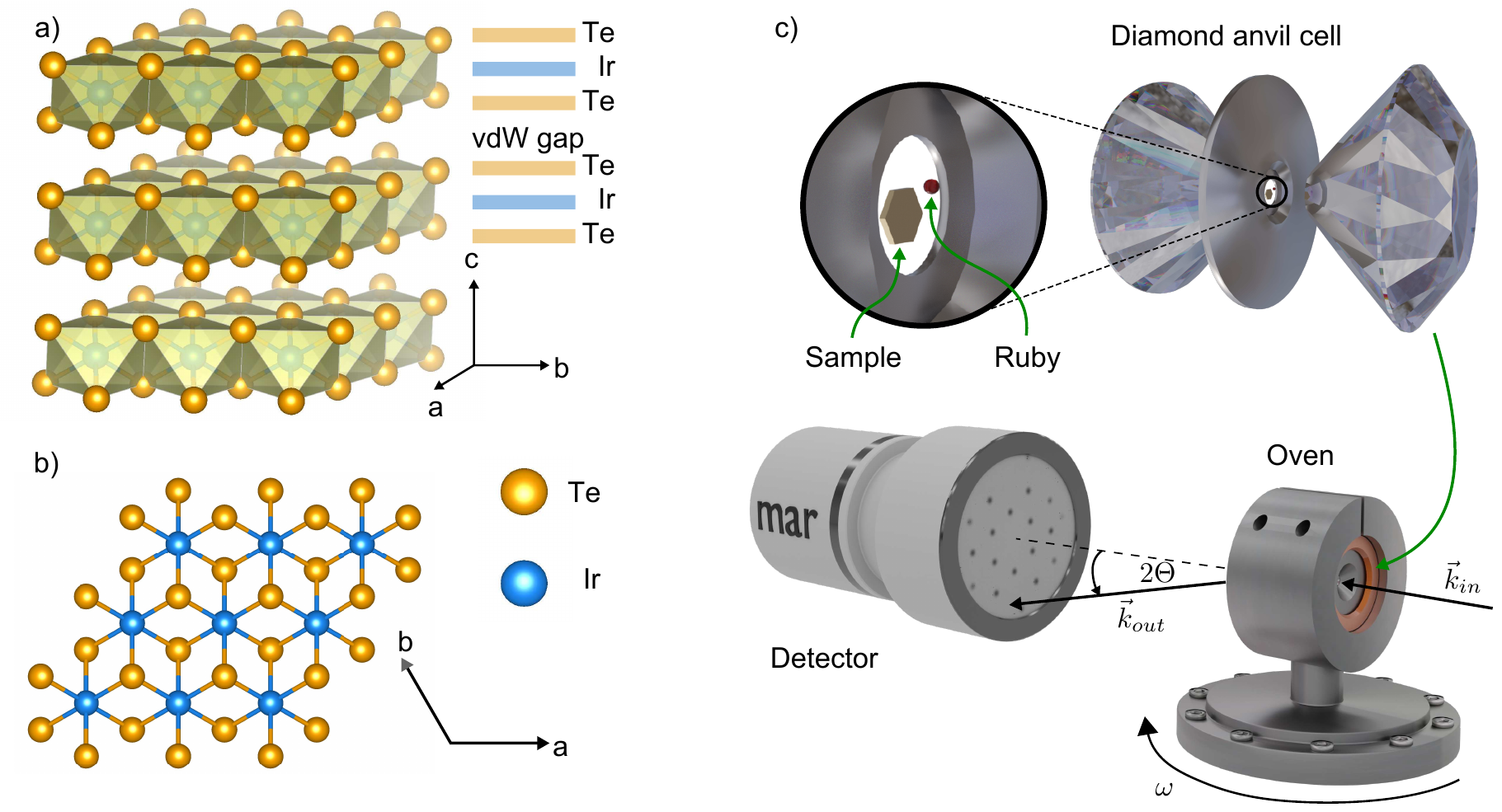}
    \caption{Trigonal 1T (spacegroup: 164) host crystal structure of \irte (a,b) and schematic of the experimental high-pressure XRD setup at the beamline ID27 of the ESRF (c).}
    \label{introfig}
\end{figure*}

The TMD material \irte is no different in this respect and, indeed, turns out to be an extremely interesting case: 
In its trigonal $1T$-structure (cf. Fig.~\ref{introfig}~(a,b)) \irte exhibits bulk Dirac points, spin-orbit driven gap inversions and the corresponding topological surface states, which all emerge from the Te $5p$ derived bands\cite{Bahramy:2018v, Nicholson:2021h,FeiBo2018,JiangLee2020}. It is particularly exciting that this system not only hosts a topological electronic band structure but also superconductivity below 3\,K, raising great hopes for intrinsic topological superconductivity~\cite{Kiswandhi2013,KudoKobayashi2013,Kamitani2013,OhYangHoribe2013}. 

But apart from this, there is also electronic order. In fact, various ordered states have been reported, which, referring to the trigonal space group $P\overline{3}m1$ (\#164, CdI$_2$-structure), can be characterized by wave vectors $\vec{q}_n=(1/(3n+2),0,1/(3n+2))$ with $n=1,2,\dots,\infty$ where $\vec{q}_\infty = (1/6,0,1/6)$. The orderings can be constructed from simple basic building blocks\cite{PascutBirol2014} containing subunits with very short Ir-Ir distances. This Ir-pair is part of a larger object, which is commonly referred to as dimer in the literature and which involves both Te- and Ir-states\cite{Saleh2020}. Bulk \irte at ambient pressure exhibits an ordered lattice of these dimers with $\vec{q}_{n=1}=(1/5,0,1/5)$ below 280\,K which transforms into a lattice with $\vec{q}_{n=2}=(1/8,0,1/8)$ at 180\,K upon further cooling\cite{PascutBirol2014}. Different ordered states can also coexist in different regions of the sample\cite{PascutBirol2014, Hsu2013,Chen:2017i}, although it appears that with decreasing temperature T larger $n$ become stable.

In bulk crystals, the ordering of dimers competes with superconductivity. As soon as this preemptive static order is suppressed, superconductivity can emerge in bulk crystals at low temperatures. In case of \irte-nanoflakes that are only a few atomic layers thick, however, the situation has been found to be different. As reported in a very recent study, the superlattice of dimers in these nanoflakes does actually support two-dimensional superconductivity~\cite{Park:2021f}. Superconductivity and ordered dimers are therefore not just competing but interacting in a more complex manner. 
One corner stone for a better understanding of the interplay between dimers, topological electrons and superconductivity in \irte is therefore an improved understanding of the dimerization process in \irte. Especially since this will expose key interactions at play in this material. 
%

So far, various mechanisms have been introduced to explain the formation of dimers in \irte, including partial Fermi-surface nesting\cite{Yang:2012a}, an orbitally induced Peierls effect\cite{Ootsuki2012}, multicenter bonds of Ir and Te\cite{Saleh2020}, the polymerization-depolymerization of Te-bonds\cite{OhYangHoribe2013} as well as electronic instabilities caused by a van Hove singularity at the Fermi-level\cite{Qian2014}. 
The multitude of scenarios quoted above demonstrates that, despite extensive research efforts, important aspects of the dimerization in \irte remain controversial. To address this issue and to shed light onto the physics at work, we performed high-pressure x-ray diffraction studies of pristine \irte single crystals. In this way, we obtain unprecedented structural information of great detail, which we then use for a thorough analysis within density functional theory. The key result of this combined study is a dramatic change in the Ir-Te-Ir bond angle within the dimers, which is instrumental for their stabilization by the formation of localized valence bonds.

\section{Results}
\subsection*{High-pressure x-ray diffraction}\label{sec:xrd}

Especially, when dealing with competing phases, as is the case here, external pressure is a particularly valuable tool, because it is a clean control parameter that avoids unwanted side effects, which may be caused by local changes induced by chemical substitution or doping. We therefore performed high-pressure XRD experiments at pressures $P$ up to 42\,GPa and temperatures between room temperature and 420\,K (see methods section). The experimental setup is sketched in Fig.~\ref{introfig}~(c). 

\begin{figure*}[t]
    \centering
    \includegraphics{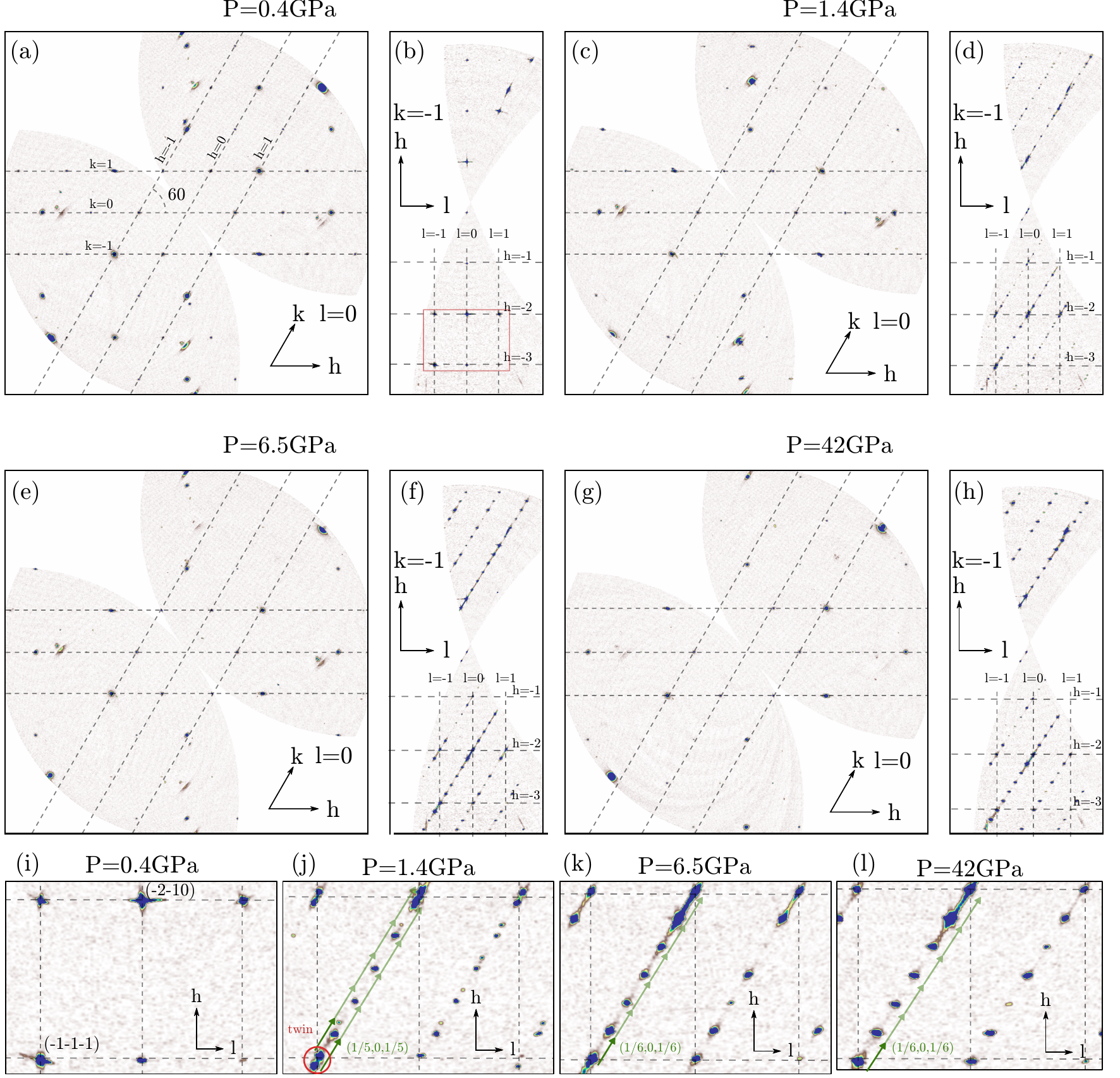}
    \caption{Reconstructed reciprocal space maps for the trigonal host phase (a,b,i) at 300\,K, the triclinic 1/5-phase (c,d,j) at 300\,K and the monoclinic 1/6-phase (g-h,k,l) at 350\,K. The pressure at which the data has been taken is indicated on top of the panels. Reciprocal lattice vectors and Miller indices $h,k$ and $l$ refer to the hexagonal cell and an approximately hexagonal cell for the 1/5- and 1/6-phase. $hk0$-planes are shown in (a,c,e,f) and $h\bar1l$-planes in (b,d,f,h). (i-l) show a magnification of the region indicated by the red rectangle in (b). Green arrows in (j-l) depict the $q$-vector. }
    \label{xrd_overview}
\end{figure*}

For each pressure point, one image has been collected by continuously recording the diffracted intensity on the detector during a large $\omega$-rotation over $\pm$ 31$^{\circ}$. These measurements provide a qualitative overview allowing quick detection of structural changes. Hence we will refer to them as overview scans in the following.
%
At selected pressure values, identified using these overview scans, single crystal XRD data sets have been collected. These data sets consist of images integrating over $\Delta\omega$ = 0.5$^{\circ}$. The obtained single crystal data has then been  used for the crystallographic structure refinements (see methods section). Detailed results of the refinements are given in the supplementary information. 

In Fig.~\ref{xrd_overview} we present representative reciprocal space maps of the data at different pressure-temperature points. At small pressures, here $P=0.4$~GPa in panels a),b) and i), we observe the trigonal phase ($P\overline{3}m1$) as expected. Fig.\,\ref{xrd_overview}\,a) show a $(h,k,0)$-plane of the reciprocal lattice, while a $(h,-1,l)$-plane is presented in Fig.\,\ref{xrd_overview}\,b,i). The vertices of the dashed lines in the panels indicate the positions of the reciprocal lattice of the trigonal phase, which coincide perfectly with all reflections originating from the sample. The additional spots outside the trigonal grid are all due to the diamond anvils of the DAC.

\begin{figure*}[t]
    \includegraphics[angle=0,width=0.75\textwidth]{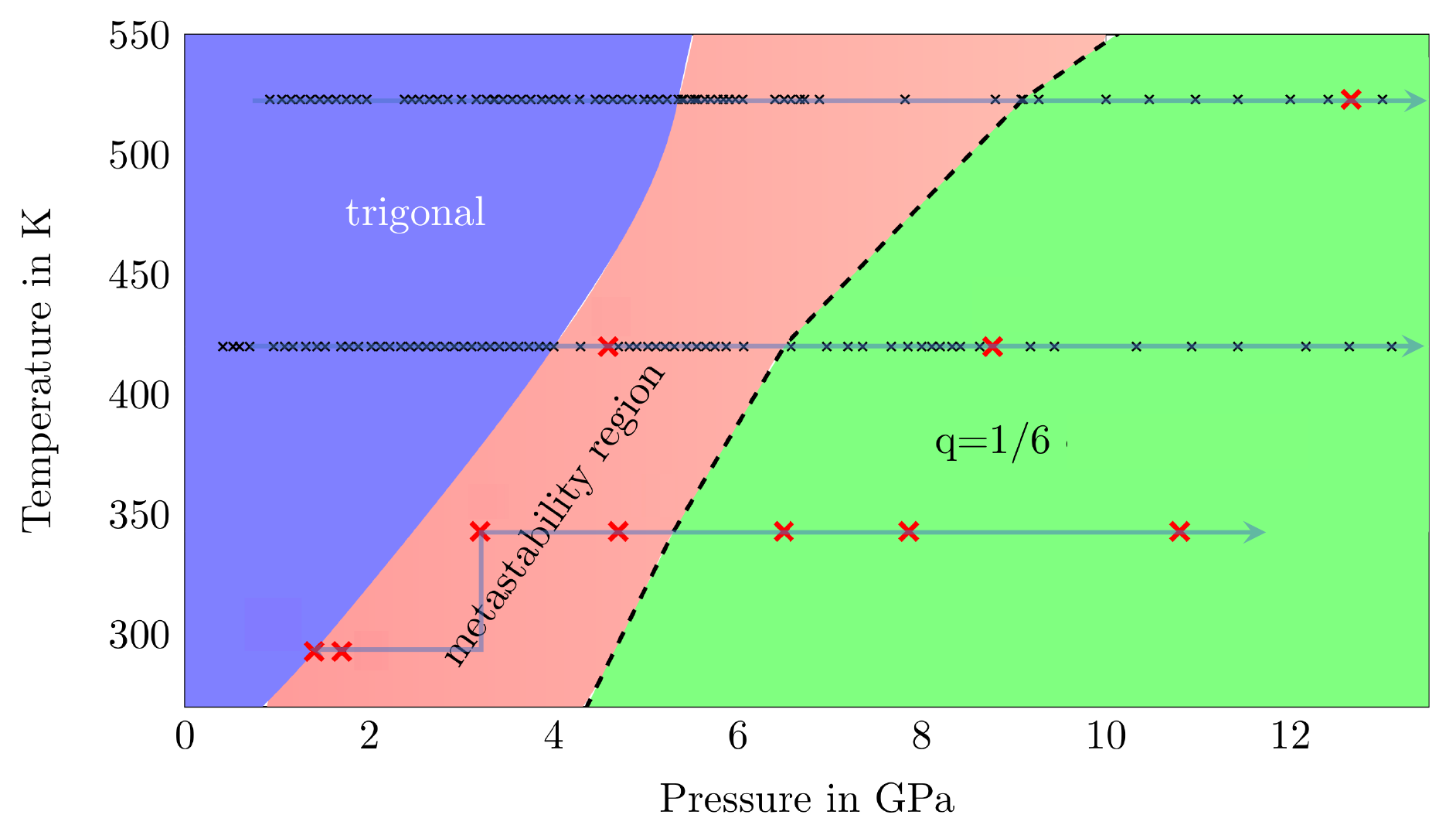}
    \caption{Pressure-temperature phase diagram of \irte as deduced from the present XRD-data. Starting from the trigonal phase without dimers, a metastability region is entered upon increasing pressure. In this region different ordered phases appear and sometimes coexist. With further increasing pressure, the long ranged ordered 1/6-phase is reached, which remains stable up to 42 GPa.
    Small (black) crosses indicate points where an overview scan has been taken. Large (red) crosses indicate points where single crystal diffraction data sets have been recorded (further explanations in the text).}
    \label{fig:phasediagram}
\end{figure*}

Upon increasing the pressure to $P=1.4$~GPa, the diffraction pattern changes drastically, as can be observed in panels c),d) and j) of Fig.\,\ref{xrd_overview}. Most prominently, additional reflections appear in the $(h,-1,l)$-plane (cf. Fig.\,\ref{xrd_overview}\,d,j)). The width and intensity of these additional reflections is comparable to that of the Bragg-reflections in the trigonal phase, i.e. we are {\it not} dealing with a small lattice modulation as, for example, in a typical weak coupling charge density wave phase. 
As can nicely be observed in Fig.\,\ref{xrd_overview}\,c) by direct inspection, the Bragg-spots at $P=1.4$~GPa deviate slightly from the trigonal reciprocal lattice (dashed lines). In fact, we find a triclinic lattice symmetry under the present conditions ($P\overline{1}$).
Nonetheless, still referring to the trigonal low-pressure phase, this new structure can be characterized by the wave vector $\vec{q}_{1}=(1/5,0,1/5)$, as it is commonly done in the literature. We will refer to this phase as the 1/5-phase in the following. 
Note that the peak splitting --particularly apparent in Fig.\,\ref{xrd_overview}\,i)-- is not caused by a phase coexistence, but is due to the twinning of the sample under the present conditions. The sample therefore shows a pure 1/5-phase, which enabled us to fully refine its structure based on the present data set (cf. supplementary information).

Upon increasing $P$ further, the 1/6-phase characterized by  $\vec{q}_{\infty}=(1/6,0,1/6)$ is entered. This is shown by the data sets for $P=6.5$\,GPa and 42\,GPa presented in Fig.\,\ref{xrd_overview}. This phase is found to be monoclinic ($C2/c$). The 1/6-phase has been found to remain stable up to the highest pressures. No significant change in position or width of the detected peaks, other than due to the compression of the lattice, has been observed up to 42\,GPa. Interestingly, the probed sample volume in the 1/6-phase was found not to be twinned, as can be observed in Fig.\,\ref{xrd_overview}. A possible reason for this detwinning is a small pressure gradient within the DAC. Also the intensities of the 1/6-phase could be refined successfully yielding a full determination of this phase (cf. supplementary information).

The measurements of our HP-XRD experiments are summarized in the $pT$ phase diagram displayed in Fig.\,\ref{fig:phasediagram}. In the region between the trigonal and the 1/6-phase, we observed the 1/5-, 1/8- and the 1/11-phase ($n=1$, 2 and 3) as well as a coexistence of some of these. For this reason it is denoted as "metastability region". 

Our refined 1/5- and 1/6-structures at $p<6.5$\,GPa resemble very closely the corresponding structures found in Se-substituted \irte at ambient pressure\cite{PascutBirol2014,PascutHaule2014}, where the concentration of dimers increases with $n$: for the 1/5-phase ($n=1$) the fraction of dimerized Ir is 2/5, for the 1/8-phase ($n=2$) it is already 1/2 and for the 1/6 ($n=\infty$) it becomes 2/3.

\begin{figure}[t!]
    \centering
    \includegraphics{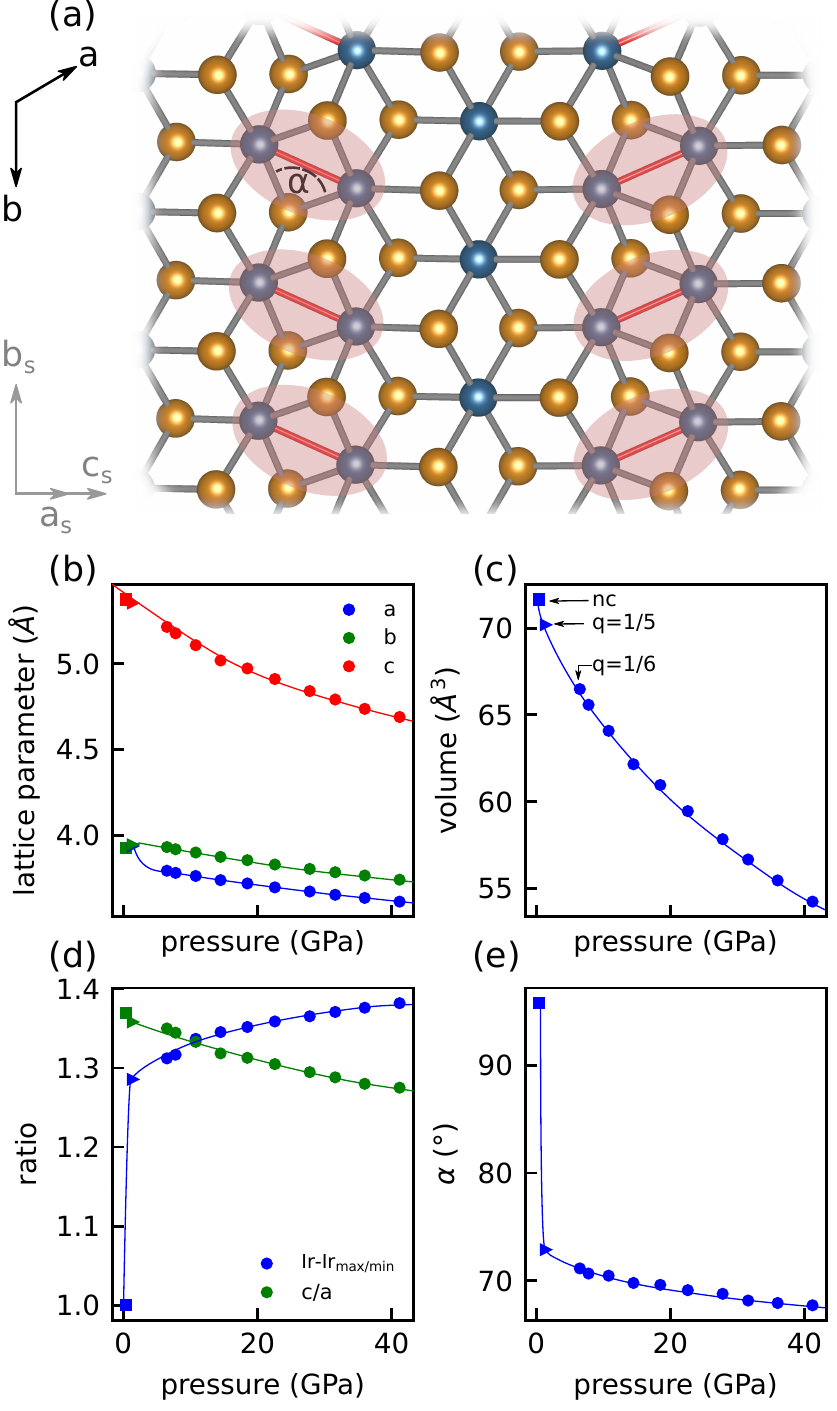}
    \caption{Structural parameters as determined from the single crystal data as a function of pressure. (a): Sketch of one Te-Ir-Te layer in the $q=1/6$ structure. The short Ir-Ir distances inside the dimers are highlighted in red. All parameters in (b-e) are derived from a transformation of the supercell back to an approximately trigonal cell. The corresponding (pseudo)-trigonal in-plane lattice vectors $\mathbf a$ and $\mathbf b$ are indicated as black arrows. The lattice vectors corresponding to the monoclinic $q=1/6$ structure ($\mathbf{a_s, b_s}$ and $\mathbf{c_s}$) are shown in gray. Note that the in-plane directions in terms of the monoclinic cell are given by $\mathbf{a_s} + \mathbf{c_s}$ and $\mathbf{b_s}$.   (b,c): The (pseudo)-trigonal lattice parameters and cell volume as a function of pressure. (d): $c/a$-ratio (green) and ratio between long and short Ir-Ir bonds (blue). (e): Ir-Te-Ir angle $\alpha$ as indicated in (a). Square, triangular and circular symbols correspond to data from the trigonal, triclinic $q=1/5$- and monoclinic $q=1/6$-phase. Solid lines are guides to the eye.}  
    \label{fig:lattice}
\end{figure}

The evolution of the lattice structure with increasing $P$ up to 42\,GPa is illustrated in Fig.\,\ref{fig:lattice}. As described above, the dimerized phases are no longer trigonal but mono- or triclinic. However, in order to facilitate the comparison among the different phases, these lower symmetry structures have been mapped onto an approximate trigonal cell. Selected structural parameters obtained by this mapping are shown in Fig.\,\ref{fig:lattice}. While in the truly trigonal phase $a=b$, for the structures of the dimerized phase only $a\simeq b$ is fulfilled. This can be observed in Fig.\,\ref{fig:lattice}\,b), where $a$ and $b$ split up upon entering the dimerized phases with increasing $P$. Panel b) and the $c/a$-ratio in panel d) show that, as expected, the $c$-axis is much more compressible than the $a$- and $b$-axis. More specifically, we obtain $4.7\times 10^{-3}\mathrm{GPa}^{-1}\simeq \kappa_c > \kappa_{a,b}\simeq 1.3 \times 10^{-3}\,\mathrm{GPa}^{-1}$ for the axial compressibilities. At 42\,GPa the $c$-axis and the unit cell volume are reduced by about 12\% and 25\%, respectively.

In the trigonal phase, each Ir-site has 6 nearest neighbor Ir-sites at the same distance $d \simeq 3.92$\,\AA within the \irte-layer. In the dimerized phases this changes and every Ir in a dimer has 1 short intra-dimer distance $d_s$ and 5 longer nearest neighbor distances of average length $d_l$. The formation of the dimers can therefore be tracked very nicely by the ratio $\gamma=d_l/d_s$, which is presented in Fig.\,\ref{fig:lattice}\,c). With increasing $P$, this ratio jumps from $\gamma=1$ in the trigonal phase (no dimers, $d=d_s=d_l$) to $\gamma \simeq 1.27$ ($d_s\simeq 3.11$\,\AA, $d_l\simeq 3.96$\,\AA) in the 1/5-phase. Since $\gamma$ can be interpreted as an order parameter of the local dimerization, its large value implies a very strong dimerization in \irte. Furthermore, the monotonous increase of $\gamma$ with $P$ shows that the dimers are more and more stabilized upon pressurization.

A key result of the present refinements concerns the Ir-Te-Ir bond angle $\alpha$ (cf. Fig.\,\ref{fig:lattice}\,a)): Across the $P$-driven transition into the 1/5-phase, the angle $\alpha$ inside the dimers 
%
%
collapses from 95.8\,$^{\circ}$ to 73.0\,$^{\circ}$, as shown in Fig.\,\ref{fig:lattice}\,e).
Note, the evolution of $\alpha$ with further increasing $P$ is smooth and monotonous also across the transition into the 1/6-phase. The reduced value of $\alpha$ is not significantly influenced by the specific long range order, but rather appears to be a characteristic property of the local dimers.

Compared to $\alpha$, the nearest neighbor Ir-Te distances change much more moderately: The single Ir-Te distance of $2.645$\,\AA\/ in the trigonal phase splits up upon entering the dimerized phases (see supplementary). This splitting is of the order of $\pm0.06$\,\AA, which amounts to about only 2\% of $2.645$\,\AA.
Also note that the Ir-Te distances of dimerized and non-dimerized Ir-sites exhibit a very similar rate of compression upon increasing $P$. 

The above analysis identifies a key structural parameter for the dimerization, which escaped attention so far: the Ir-Te-Ir bond angle $\alpha$. The question now obviously is, which role $\alpha$ plays for the stabilization of dimers in \irte.

\begin{figure*}[t]
   \includegraphics[width=\textwidth]{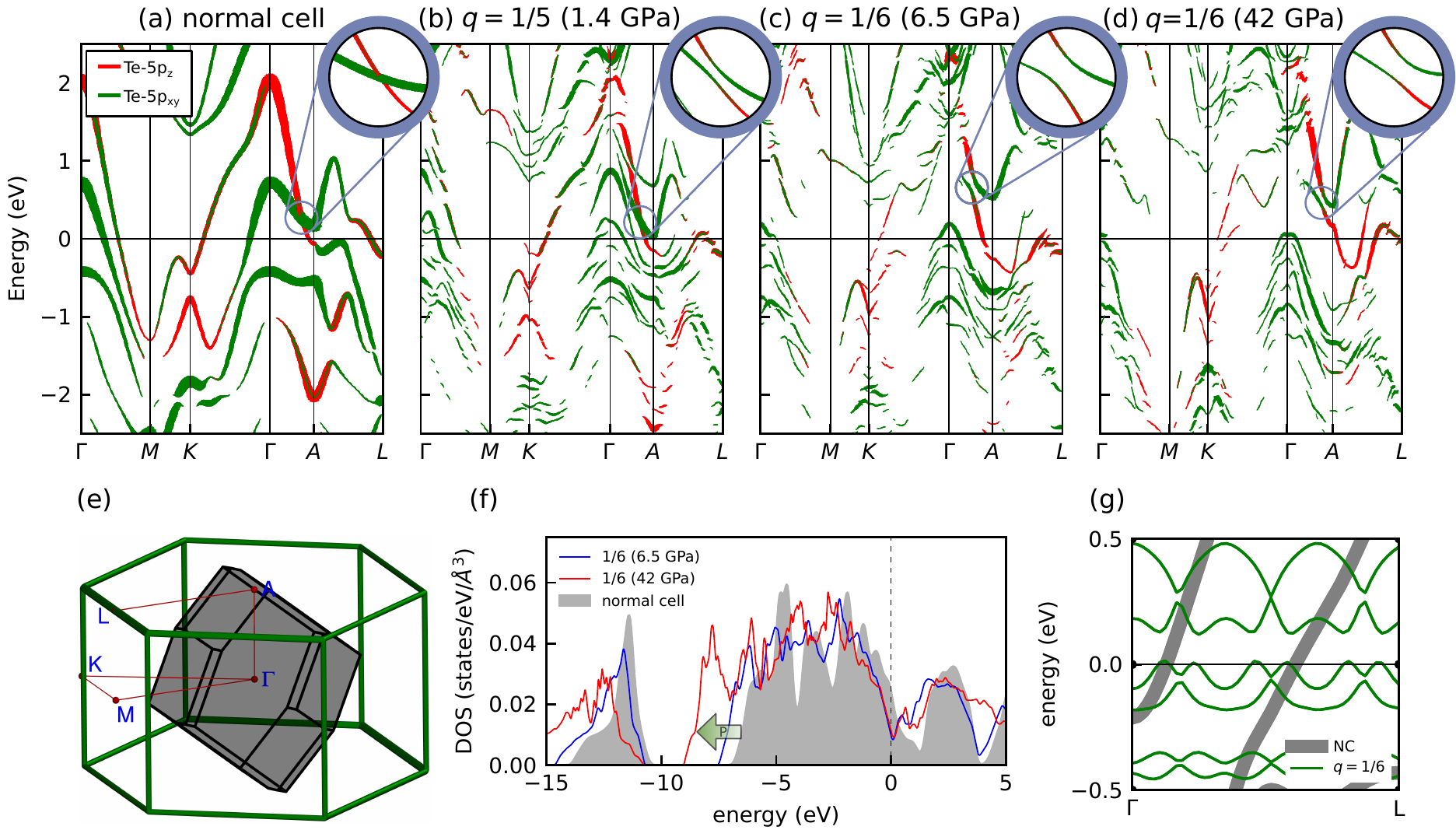}
    \caption{Electronic structure as a function of hydrostatic pressure. (a-d): Fat bands representing the Te-5p character for the  trigonal structure at ambient conditions (a), the $q=1/5$ structure (b) and the $q=1/6$ structure (c,d). Band weights of the $q=1/5$ and $q=1/6$ phase were unfolded into the Brillouin zone of the trigonal structure. A magnification of the regions around the BDP is shown in the top right corner for each panel. The high symmetry points refer to the trigonal Brillouin zone (BZ) shown in green in (e). The BZ of the monoclinic 1/6-phase is illustrated in gray in (e). (f): Total density of states (DOS) at ambient conditions (normal cell) and for the 1/6-phase at 6.5 GPa. The green arrow indicates the shift of the dimer bonding states to lower energies with increasing pressure. (g): Comparison of the band structure along the $q$-direction of the 1/6-phase with the normal cell band structure.}
    \label{fig:dft_electrons}
\end{figure*}

\subsection*{Density functional theory}\label{sec:dft}

To address precisely this question and to elucidate the relation between the formation of dimers and the electronic structure, we performed density functional theory (DFT) calculations based on the experimentally determined lattice structures.
Calculations for 4 different structures were performed, namely (i) the trigonal structure, (ii) the pure 1/5-phase, (iii) the 1/6-phase at 6.5\,GPa and (iv) the 1/6-phase at 42\,GPa (cf. structural parameters in the supplementary information). The electronic bands obtained for the 1/5- and 1/6-phase have been unfolded onto the original Brillouin zone of the trigonal structure, using the approach described in Ref.~\onlinecite{KuPRL2010}. In this way the electronic bands for the three different structures can be compared directly. 


We first discuss the evolution of the bulk DFT band structure with increasing hydrostatic pressure. As described very nicely in Ref.\,\onlinecite{Bahramy:2018v}, the chalcogenide p-orbital manifold is the most important one for the formation of Dirac-cones, topological surface states and topological surface resonances. In Fig.\,\ref{fig:dft_electrons}\,a)--d) we therefore show so-called fat bands in which the Te $5p_{x,y,z}$-character is represented by the line thickness and the colour. 

The results for the trigonal phase in Fig.\,\ref{fig:dft_electrons}\,a) agree very well with previously published DFT- and ARPES-results\cite{Qian2014}, including the van-Hove singularity at the Fermi-level along the $AL$-direction.
Our DFT band structure also exhibits precisely the features that have been discussed in Ref.\,\onlinecite{Bahramy:2018v}: Firstly, there is a type II bulk Dirac point (BDP) along the $\Gamma A$-direction, which is protected by the $C_{3v}$-symmetry of the lattice. More to the point, the $C_{3v}$-symmetry forbids hybridization of Te $5p_{x,y}$ and $5p_{z}$ bands along the $\Gamma A$-direction so that there is no hybridization gap. The two bands therefore cross at the BDP, forming states that correspond to massless Dirac-fermions. In the present case, however, the BDP is above the Fermi-level. Secondly, inverted band gaps of the bulk electronic structure are obtained as well. These inverted band gaps are expected to result in topological surface states and resonances\,\cite{Bahramy:2018v}, which indeed appears to be consistent with a very recent ARPES-study\,\cite{Nicholson:2021h}.

\begin{figure*}[t!]
    \centering
    \includegraphics{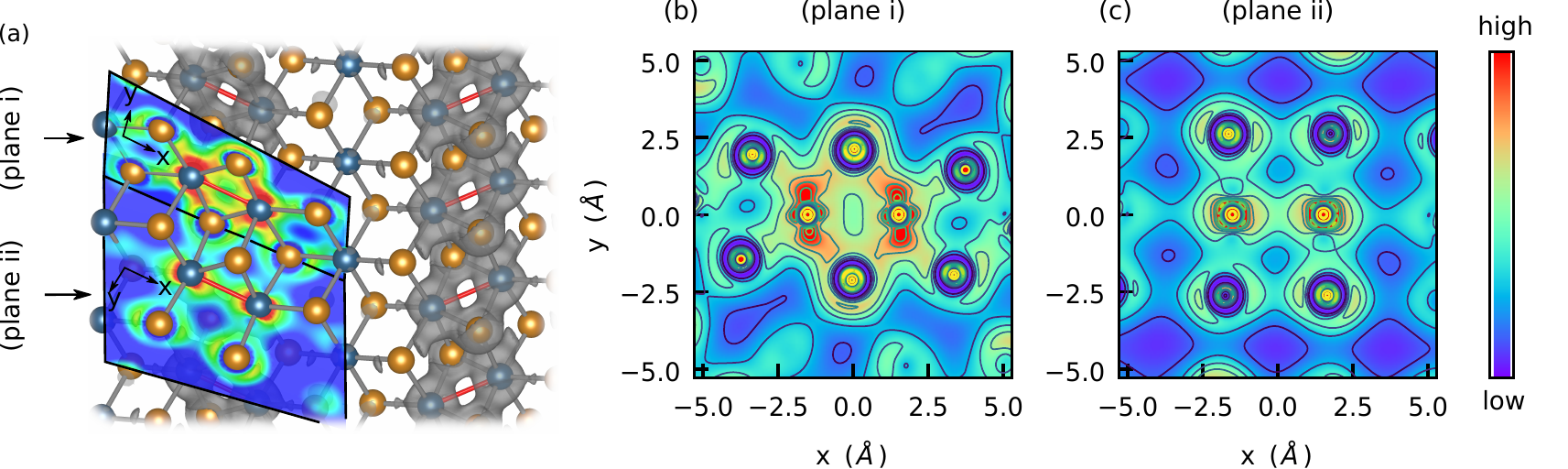}
    \caption{Charge density distribution of the bonding dimer states for the 1/6-phase at 6.5\,GPa (further explanations in the text). Left: Cuts through the charge density parallel to two different planes placed into the lattice structure. The short Ir-Ir distances within a dimer are indicated by red lines. Gray shaded ring-like structures indicate the charge distribution of the bonding states of the dimers. Right: Contour plots of the charge densities of these covalent bonds for plane i and plane ii. Position of the planes is indicated in the left panel.}
    \label{fig:densities}
\end{figure*}

Using the structural refinements described above as input for DFT, we can now address the evolution of the electronic structure with increasing hydrostatic pressure. As is obvious from Figs.\,\ref{fig:dft_electrons}\,b)--d), the pressure-induced appearance of dimers causes drastic changes in the electronic band structure: Starting from the triclinic 1/5-phase, a strong splitting of the bands is apparent with states entering the energy regions of the inverted band gaps. The band crossing of the BDP is lost and a gap opens, which corresponds to the formation of massive Dirac states, due to the broken $C_{3v}$-symmetry. Upon increasing the applied pressure further, the monoclinic 1/6-phase occurs. In this phase the band gap at the former BDP opens up further.

By comparing the energies of the $5p_z$-states at $\Gamma$ and $A$, one can see that the energy spread of these states increases with increasing pressure as expected from the reduced $c/a$-ratio. But also the $5p_{xy}$-type bands change strongly. In particular we also find that the van-Hove singularity along the $AL$-direction is pushed further below the Fermi-level in agreement with in Ref.\,\onlinecite{Qian2014}.

Along the $\Gamma L$-direction, which is parallel to the the wave vectors of the ordered dimers, a gap opens between the occupied and the unoccupied states (Fig.\,\ref{fig:dft_electrons}\,g))
This agrees perfectly with the earlier conclusion that --in direct space-- the electrons become confined to the two-dimensional layers between neighbouring dimer-walls\,\cite{PascutBirol2014,Park:2021f}.

Fig.\,\ref{fig:dft_electrons}\,f) shows the evolution of the density of states (DOS) with pressure. Also here the drastic changes in the electronic structure can clearly be discerned. Indeed, the changes caused by applying $P$ correspond closely to what has been found for the 1/5-phase at ambient pressure but lower temperature\cite{Saleh2020}.
Most importantly for the following discussion are the bonding states formed by Te $5p$ and Ir $5d$ that are pushed out of the broad band continuum between 0\,eV and -6\,eV, once the dimers form. 
%
Using the results from our XRD-measurements, we can now follow the evolution of these bonding dimer-states with increasing $P$. As can be observed in Fig.\,\ref{fig:dft_electrons}\,f), these bonding dimer-states shift very strongly to lower and lower energies with $P$ growing from 1.4\,GPa to 41\,GPa. Indeed there is a huge shift of about 2\,eV, which amounts to a relative change of approximately 30\,\%. 
%

The energy resolved charge distribution of these bonding dimer-states is illustrated in Fig.\,\ref{fig:densities}. As before, the short Ir-Ir distances within the dimers in panel a) are indicated by red lines. Inspection of Fig.\,\ref{fig:densities}\,a) reveals a ring-shaped charge-density distribution going {\it around} every short Ir-Ir link (gray colored charge clouds which correspond to an isolevel of $6\times10^{-4}$ e/\AA$^{-3}$). This can be also seen in the cuts through the charge density along the planes i) and ii) in Fig.\,\ref{fig:densities}\,a). The same contour plots are also shown in  more detail in Figs.\,\ref{fig:densities}\,b) and c). Note the strong distortion of the angle $\alpha$ of the Ir-Te-Ir links that is clearly apparent in Figs.\,\ref{fig:densities}\,a,b).

The most bonding states of the dimer are therefore {\it not} due to direct Ir-Ir interactions. A detailed analysis of the chemical bonds in \irte indeed showed that direct Ir-Ir bonding is not relevant\,\cite{Saleh2020}. Instead we find that the ring-shaped covalent bond, which is formed by Ir $5d$ and Te $5p$ states, plays a key role for the stabilization of the dimers. This is verified by the strong $P$-driven stabilization of this bond, which goes hand in hand with the experimentally observed change in $\alpha$.

\section{Discussion}

The important role of the ring-shaped covalent bonds for the dimers has also been found in Ref.\,\onlinecite{Saleh2020}. The key new result of the present study concerns the stabilization of these bonds: 
Our XRD data reveals that the bond angle $\alpha$ changes drastically with pressure, while the Ir-Te distances change only slightly and continuously across the phase transitions. At the same time, our DFT results show that with increasing $P$ the localized covalent bonds are pushed out of the broad band continuum formed by itinerant Ir $5d$ and Te $5p$ states. The very large energy shifts of up to 2\,eV cannot be explained by the small changes of the Ir-Te bond lengths, which, furthermore, change by essentially the same amount for Ir-sites within and outside a dimer. 
%
%
We therefore conclude that $\alpha$ is the essential structural degree of freedom for the stabilization of the ring-shapes covalent bond, which correspond to a valence bond on two edge-sharing IrTe$_6$-octahedra formed by Te $5p$- and Ir $5d$-states.

One way to rationalize this, is to consider two Wannier-orbitals of $e_g$-symmetry on each of the two edge-sharing IrTe$_6$-octahedra of a dimer. In the edge-sharing geometry, the overlap of these two orbitals changes very strongly with $\alpha$. More specifically, the overlap increases rapidly as $\alpha$ deviates from $90^\circ$. As a result, the observed reduction of $\alpha$ towards $70^\circ$ with increasing $P$ yields a strong stabilization of the valence bond between the two IrTe$_6$-octahedra. 

There are two key features of this mechanism: First, it is essentially local, since it corresponds to the formation of a valence bond 
%
centered on two edge-sharing IrTe$_6$-octahedra. Importantly, this valence bond is not stabilized by direct Ir-Ir interactions, which is very surprising considering strong shortening of the Ir-Ir distance within a dimer. Second, the strong change of the electronic binding energy with $\alpha$ implies a very strong  coupling of electrons and phonons. The large structural distortions observed experimentally directly reflect this strong electron-phonon coupling.

The mechanism identified in this study is different from other mechanisms discussed previously, especially those referring to weak coupling scenarios based on the electronic band structure. Such scenarios are not easily reconciled with the various $\vec{q}_n$ observed experimentally, because they involve specific wave vectors defined by the band structure of the trigonal phase. The local $\alpha$-dependent hybridization, however, is not tied to details of the band structure, i.e. it can easily accommodate the different and often coexisting $\vec{q}_n$ found in experiment. We emphasize that the specific type of long-range order under certain conditions, i.e. the specific $\vec{q}_n$, may very well be determined by the underlying electronic band structure. According to our analysis, however, the primary stabilization mechanism of the dimers is the local $\alpha$-dependent hybridization between neighboring IrTe$_6$-octahedra.


The formation of localized valence bonds will compete against the delocalization of charges in band-like states. In other words, the local stabilization mechanism must be strong enough to pull the bonding states out of the broad band continuum. According to DFT this is indeed the case here. Nonetheless, this competition may explain why so far no phase, in which all Ir-sites are part of a valence bond --or a dimer for that matter-- have been observed and all valence bond crystals still feature mobile charge carriers.

We also note that the valence bond formation described here does not involve charge order in the sense of Ir-sites with significantly different valences. In the present multi-site situation with extended Ir- and Te-states it is hardly possible to assign a certain charge to a specific lattice site. In addition we are dealing with very similar electron affinities of Ir and Te. A charge ordering scenario therefore appears to be inadequate.  Notwithstanding, the valence bond crystals of course posses inequivalent Ir-sites which, for instance, explain the peak splittings observed in x-ray photo emission spectroscopy\cite{Nicholson:2021h} -- even without different valences of Ir. 

The formation of the valence bonds involves changes in the occupied Te- and Ir-states, which is at least qualitatively consistent with the depolymerization scenarios discussed earlier\cite{OhYangHoribe2013}. Referring to this scenario, our analysis provides a mechanism underlying the depolymerization: The formation of strong intra-layer valence bonds on neighboring edge-sharing IrTe$_6$-octahedra results in a weakening of the inter-layer Te-Te bonds.



Since the stabilizing mechanism of the valence bonds is essentially local, it does not depend on details of the band structure and should be weakly dependent on doping. This mechanism can therefore be expected to be active in doped \irte as well. It is hence very interesting to ask whether local and possibly dynamic valence bonds also exist in the trigonal phases and whether they could play a role for the superconductivity in doped \irte. Going beyond the specific case of \irte, it will now also be important to investigate the relevance of this mechanism in other materials featuring extended $p$ and $d$ valence states, especially with regard to magnetic instabilities, topology and superconducting states.
These questions certainly deserve to by scrutinized in future studies.


\section{Methods}
{\it High-pressure X-ray diffraction:} The high-pressure XRD measurements have been performed at the beamline ID27 of the European Synchrotron Radiation Facility (ESRF) in Grenoble, using a monochromatic beam with a photon energy of 33 keV (\mbox{$\lambda$=0.3738 \AA}) and a spot size of 3x3 $\mu$m. High-quality single crystals have been loaded into a membrane-driven diamond anvil cell (DAC) with helium as pressure transmitting medium. The pressure inside the DAC has been monitored {\it in-situ} using the R$_{1,2}$ fluorescence of Cr-centers in a ruby sphere close to the sample. 
For the present temperature dependent measurements between room temperature and 420\,K, the DAC has been installed in an oven, which was itself mounted on a single axis ($\omega$) goniometer that has been equipped with a MARCCD area detector for efficient detection of the diffracted intensities.


{\it X-ray data analysis and structure refinement:} All data processing was performed in the CrysAlisPro software suite (version 171.39.46)~\cite{CrysAlis}. Empirical absorption correction was applied using spherical harmonics, implemented in SCALE3 ABSPACK scaling algorithm. The subsequent structure solution and weighted fullmatrix least-squares refinement on F$^2$ were done with SHELXT-2014/5 (Ref.~\onlinecite{Sheldrick2015a}) and SHELXL-2018/3 (Ref.~\onlinecite{Sheldrick2015}) as implemented in the WinGx 2018.3 program suite (Ref.~\onlinecite{Farrugia2012}). As reciprocal space coverage is significantly reduced by the DAC, the crystal structures were exclusively refined with isotropic displacement parameters. The ratio of independent reflections to refined parameters exceeds a value of ten for all data sets, with the exception of the data set for the trigonal phase measured at $P=0.4$~GPa. The parameters of the data collection and the results of the structural refinement are summarized in Supplementary Table. The atomic positions and isotropic displacement parameters are listed in Supplementary Table.

{\it Density functional Theory:} The first-principle calculations have been done using the FPLO package (version 18)~\cite{Koepernik1999}.
We used the local density approximation (LDA) and the generalized gradient approximation (GGA) of the exchange-correlation potential as parameterized by Perdew, Wang (Ref.~\onlinecite{Perdew1992}) and Perdew, Burke and Ernzerhof (Ref.~\onlinecite{PerdewBurkeErnzerhof1996}), respectively. We found very little differences in the resulting band structure for the two approximations. Therefore we present the results from the calculations performed within LDA.  
In order to account for spin-orbit coupling the calculations were performed within the full 4-component Dirac-Kohn-Sham
theory as implemented in FPLO~\cite{Eschrig2004}. The total density was converged on a grid of $24\times24\times24$ and $12\times12\times12$ irreducible k-points for the normal cell and the super cell calculations, respectively. Brillouin zone integration was done using the tetrahedron method.

\section*{Acknowledgements}
This research has been supported by the Deutsche Forschungsgemeinschaft through the the projects C06 and C09 of the SFB 1143 (project-id 247310070) and the W\"urzburg-Dresden Cluster of Excellence on Complexity and Topology in Quantum Matter–ct.qmat (EXC 2147, project-id 390858490). The work at Rutgers University was supported by the DOE under Grant No. DOE: DE-FG02-07ER46382.

\section*{Author contributions}
S.-W.C. and J.G conceived the high-pressure diffraction experiment. S.-W.C. and J.Y provided \irte single crystals. T.R, J.T., M.K., G.G., V.S. and JG conducted the diffraction experiment at the ESRF. T.R., Q.S., M.K. and J.G. analysed the diffraction data, T.R. performed the DFT-study. The manuscript has been written with contributions from all authors.  

%





\begin{thebibliography}{34}%
\makeatletter
\providecommand \@ifxundefined [1]{%
 \@ifx{#1\undefined}
}%
\providecommand \@ifnum [1]{%
 \ifnum #1\expandafter \@firstoftwo
 \else \expandafter \@secondoftwo
 \fi
}%
\providecommand \@ifx [1]{%
 \ifx #1\expandafter \@firstoftwo
 \else \expandafter \@secondoftwo
 \fi
}%
\providecommand \natexlab [1]{#1}%
\providecommand \enquote  [1]{``#1''}%
\providecommand \bibnamefont  [1]{#1}%
\providecommand \bibfnamefont [1]{#1}%
\providecommand \citenamefont [1]{#1}%
\providecommand \href@noop [0]{\@secondoftwo}%
\providecommand \href [0]{\begingroup \@sanitize@url \@href}%
\providecommand \@href[1]{\@@startlink{#1}\@@href}%
\providecommand \@@href[1]{\endgroup#1\@@endlink}%
\providecommand \@sanitize@url [0]{\catcode `\\12\catcode `\$12\catcode
  `\&12\catcode `\#12\catcode `\^12\catcode `\_12\catcode `\%12\relax}%
\providecommand \@@startlink[1]{}%
\providecommand \@@endlink[0]{}%
\providecommand \url  [0]{\begingroup\@sanitize@url \@url }%
\providecommand \@url [1]{\endgroup\@href {#1}{\urlprefix }}%
\providecommand \urlprefix  [0]{URL }%
\providecommand \Eprint [0]{\href }%
\providecommand \doibase [0]{http://dx.doi.org/}%
\providecommand \selectlanguage [0]{\@gobble}%
\providecommand \bibinfo  [0]{\@secondoftwo}%
\providecommand \bibfield  [0]{\@secondoftwo}%
\providecommand \translation [1]{[#1]}%
\providecommand \BibitemOpen [0]{}%
\providecommand \bibitemStop [0]{}%
\providecommand \bibitemNoStop [0]{.\EOS\space}%
\providecommand \EOS [0]{\spacefactor3000\relax}%
\providecommand \BibitemShut  [1]{\csname bibitem#1\endcsname}%
\let\auto@bib@innerbib\@empty
\bibitem [{\citenamefont {Stewart}(2017)}]{Stewart:2017x}%
  \BibitemOpen
  \bibfield  {author} {\bibinfo {author} {\bibfnamefont {G.~R.}\ \bibnamefont
  {Stewart}},\ }\href@noop {} {\bibfield  {journal} {\bibinfo  {journal}
  {Advances in Physics}\ }\textbf {\bibinfo {volume} {66}},\ \bibinfo {pages}
  {75} (\bibinfo {year} {2017})}\BibitemShut {NoStop}%
\bibitem [{\citenamefont {Arpaia}\ \emph {et~al.}(2019)\citenamefont {Arpaia},
  \citenamefont {Caprara}, \citenamefont {Fumagalli}, \citenamefont
  {De~Vecchi}, \citenamefont {Peng}, \citenamefont {Andersson}, \citenamefont
  {Betto}, \citenamefont {De~Luca}, \citenamefont {Brookes}, \citenamefont
  {Lombardi}, \citenamefont {Salluzzo}, \citenamefont {Braicovich},
  \citenamefont {Di~Castro}, \citenamefont {Grilli},\ and\ \citenamefont
  {Ghiringhelli}}]{Arpaia:2019aa}%
  \BibitemOpen
  \bibfield  {author} {\bibinfo {author} {\bibfnamefont {R.}~\bibnamefont
  {Arpaia}}, \bibinfo {author} {\bibfnamefont {S.}~\bibnamefont {Caprara}},
  \bibinfo {author} {\bibfnamefont {R.}~\bibnamefont {Fumagalli}}, \bibinfo
  {author} {\bibfnamefont {G.}~\bibnamefont {De~Vecchi}}, \bibinfo {author}
  {\bibfnamefont {Y.~Y.}\ \bibnamefont {Peng}}, \bibinfo {author}
  {\bibfnamefont {E.}~\bibnamefont {Andersson}}, \bibinfo {author}
  {\bibfnamefont {D.}~\bibnamefont {Betto}}, \bibinfo {author} {\bibfnamefont
  {G.~M.}\ \bibnamefont {De~Luca}}, \bibinfo {author} {\bibfnamefont {N.~B.}\
  \bibnamefont {Brookes}}, \bibinfo {author} {\bibfnamefont {F.}~\bibnamefont
  {Lombardi}}, \bibinfo {author} {\bibfnamefont {M.}~\bibnamefont {Salluzzo}},
  \bibinfo {author} {\bibfnamefont {L.}~\bibnamefont {Braicovich}}, \bibinfo
  {author} {\bibfnamefont {C.}~\bibnamefont {Di~Castro}}, \bibinfo {author}
  {\bibfnamefont {M.}~\bibnamefont {Grilli}}, \ and\ \bibinfo {author}
  {\bibfnamefont {G.}~\bibnamefont {Ghiringhelli}},\ }\href@noop {} {\bibfield
  {journal} {\bibinfo  {journal} {Science}\ }\textbf {\bibinfo {volume}
  {365}},\ \bibinfo {pages} {906} (\bibinfo {year} {2019})}\BibitemShut
  {NoStop}%
\bibitem [{\citenamefont {Savary}\ and\ \citenamefont
  {Balents}(2017)}]{Savary:2017a}%
  \BibitemOpen
  \bibfield  {author} {\bibinfo {author} {\bibfnamefont {L.}~\bibnamefont
  {Savary}}\ and\ \bibinfo {author} {\bibfnamefont {L.}~\bibnamefont
  {Balents}},\ }\href@noop {} {\bibfield  {journal} {\bibinfo  {journal}
  {Reports on Progress in Physics}\ }\textbf {\bibinfo {volume} {80}},\
  \bibinfo {pages} {016502} (\bibinfo {year} {2017})}\BibitemShut {NoStop}%
\bibitem [{\citenamefont {Yan}\ and\ \citenamefont {Felser}(2017)}]{Yan:2017a}%
  \BibitemOpen
  \bibfield  {author} {\bibinfo {author} {\bibfnamefont {B.}~\bibnamefont
  {Yan}}\ and\ \bibinfo {author} {\bibfnamefont {C.}~\bibnamefont {Felser}},\
  }\href@noop {} {\bibfield  {journal} {\bibinfo  {journal} {Annual Review of
  Condensed Matter Physics}\ }\textbf {\bibinfo {volume} {8}},\ \bibinfo
  {pages} {337} (\bibinfo {year} {2017})}\BibitemShut {NoStop}%
\bibitem [{\citenamefont {Wehling}\ \emph {et~al.}(2014)\citenamefont
  {Wehling}, \citenamefont {Black-Schaffer},\ and\ \citenamefont
  {Balatsky}}]{Wehling:2014m}%
  \BibitemOpen
  \bibfield  {author} {\bibinfo {author} {\bibfnamefont {T.~O.}\ \bibnamefont
  {Wehling}}, \bibinfo {author} {\bibfnamefont {A.~M.}\ \bibnamefont
  {Black-Schaffer}}, \ and\ \bibinfo {author} {\bibfnamefont {A.~V.}\
  \bibnamefont {Balatsky}},\ }\href@noop {} {\bibfield  {journal} {\bibinfo
  {journal} {Advances in Physics}\ }\textbf {\bibinfo {volume} {63}},\ \bibinfo
  {pages} {1} (\bibinfo {year} {2014})}\BibitemShut {NoStop}%
\bibitem [{\citenamefont {Rossnagel}(2011)}]{Rossnagel:2011a}%
  \BibitemOpen
  \bibfield  {author} {\bibinfo {author} {\bibfnamefont {K.}~\bibnamefont
  {Rossnagel}},\ }\href@noop {} {\bibfield  {journal} {\bibinfo  {journal}
  {Journal of Physics: Condensed Matter}\ }\textbf {\bibinfo {volume} {23}},\
  \bibinfo {pages} {213001} (\bibinfo {year} {2011})}\BibitemShut {NoStop}%
\bibitem [{\citenamefont {Yang}\ \emph {et~al.}(2017)\citenamefont {Yang},
  \citenamefont {Kim}, \citenamefont {Chhowalla},\ and\ \citenamefont
  {Lee}}]{Yang:2017e}%
  \BibitemOpen
  \bibfield  {author} {\bibinfo {author} {\bibfnamefont {H.}~\bibnamefont
  {Yang}}, \bibinfo {author} {\bibfnamefont {S.~W.}\ \bibnamefont {Kim}},
  \bibinfo {author} {\bibfnamefont {M.}~\bibnamefont {Chhowalla}}, \ and\
  \bibinfo {author} {\bibfnamefont {Y.~H.}\ \bibnamefont {Lee}},\ }\href@noop
  {} {\bibfield  {journal} {\bibinfo  {journal} {Nature Physics}\ }\textbf
  {\bibinfo {volume} {13}},\ \bibinfo {pages} {931} (\bibinfo {year}
  {2017})}\BibitemShut {NoStop}%
\bibitem [{\citenamefont {Gye}\ \emph {et~al.}(2019)\citenamefont {Gye},
  \citenamefont {Oh},\ and\ \citenamefont {Yeom}}]{Gye:2019s}%
  \BibitemOpen
  \bibfield  {author} {\bibinfo {author} {\bibfnamefont {G.}~\bibnamefont
  {Gye}}, \bibinfo {author} {\bibfnamefont {E.}~\bibnamefont {Oh}}, \ and\
  \bibinfo {author} {\bibfnamefont {H.~W.}\ \bibnamefont {Yeom}},\ }\href@noop
  {} {\bibfield  {journal} {\bibinfo  {journal} {Phys. Rev. Lett.}\ }\textbf
  {\bibinfo {volume} {122}},\ \bibinfo {pages} {016403} (\bibinfo {year}
  {2019})}\BibitemShut {NoStop}%
\bibitem [{\citenamefont {Bahramy}\ \emph {et~al.}(2018)\citenamefont
  {Bahramy}, \citenamefont {Clark}, \citenamefont {Yang}, \citenamefont {Feng},
  \citenamefont {Bawden}, \citenamefont {Riley}, \citenamefont {Markovi{\'c}},
  \citenamefont {Mazzola}, \citenamefont {Sunko}, \citenamefont {Biswas},
  \citenamefont {Cooil}, \citenamefont {Jorge}, \citenamefont {Wells},
  \citenamefont {Leandersson}, \citenamefont {Balasubramanian}, \citenamefont
  {Fujii}, \citenamefont {Vobornik}, \citenamefont {Rault}, \citenamefont
  {Kim}, \citenamefont {Hoesch}, \citenamefont {Okawa}, \citenamefont
  {Asakawa}, \citenamefont {Sasagawa}, \citenamefont {Eknapakul}, \citenamefont
  {Meevasana},\ and\ \citenamefont {King}}]{Bahramy:2018v}%
  \BibitemOpen
  \bibfield  {author} {\bibinfo {author} {\bibfnamefont {M.~S.}\ \bibnamefont
  {Bahramy}}, \bibinfo {author} {\bibfnamefont {O.~J.}\ \bibnamefont {Clark}},
  \bibinfo {author} {\bibfnamefont {B.~J.}\ \bibnamefont {Yang}}, \bibinfo
  {author} {\bibfnamefont {J.}~\bibnamefont {Feng}}, \bibinfo {author}
  {\bibfnamefont {L.}~\bibnamefont {Bawden}}, \bibinfo {author} {\bibfnamefont
  {J.~M.}\ \bibnamefont {Riley}}, \bibinfo {author} {\bibfnamefont
  {I.}~\bibnamefont {Markovi{\'c}}}, \bibinfo {author} {\bibfnamefont
  {F.}~\bibnamefont {Mazzola}}, \bibinfo {author} {\bibfnamefont
  {V.}~\bibnamefont {Sunko}}, \bibinfo {author} {\bibfnamefont
  {D.}~\bibnamefont {Biswas}}, \bibinfo {author} {\bibfnamefont {S.~P.}\
  \bibnamefont {Cooil}}, \bibinfo {author} {\bibfnamefont {M.}~\bibnamefont
  {Jorge}}, \bibinfo {author} {\bibfnamefont {J.~W.}\ \bibnamefont {Wells}},
  \bibinfo {author} {\bibfnamefont {M.}~\bibnamefont {Leandersson}}, \bibinfo
  {author} {\bibfnamefont {T.}~\bibnamefont {Balasubramanian}}, \bibinfo
  {author} {\bibfnamefont {J.}~\bibnamefont {Fujii}}, \bibinfo {author}
  {\bibfnamefont {I.}~\bibnamefont {Vobornik}}, \bibinfo {author}
  {\bibfnamefont {J.~E.}\ \bibnamefont {Rault}}, \bibinfo {author}
  {\bibfnamefont {T.~K.}\ \bibnamefont {Kim}}, \bibinfo {author} {\bibfnamefont
  {M.}~\bibnamefont {Hoesch}}, \bibinfo {author} {\bibfnamefont
  {K.}~\bibnamefont {Okawa}}, \bibinfo {author} {\bibfnamefont
  {M.}~\bibnamefont {Asakawa}}, \bibinfo {author} {\bibfnamefont
  {T.}~\bibnamefont {Sasagawa}}, \bibinfo {author} {\bibfnamefont
  {T.}~\bibnamefont {Eknapakul}}, \bibinfo {author} {\bibfnamefont
  {W.}~\bibnamefont {Meevasana}}, \ and\ \bibinfo {author} {\bibfnamefont
  {P.~D.~C.}\ \bibnamefont {King}},\ }\href@noop {} {\bibfield  {journal}
  {\bibinfo  {journal} {Nature Materials}\ }\textbf {\bibinfo {volume} {17}},\
  \bibinfo {pages} {21} (\bibinfo {year} {2018})}\BibitemShut {NoStop}%
\bibitem [{\citenamefont {Nicholson}\ \emph {et~al.}(2021)\citenamefont
  {Nicholson}, \citenamefont {Rumo}, \citenamefont {Pulkkinen}, \citenamefont
  {Kremer}, \citenamefont {Salzmann}, \citenamefont {Mottas}, \citenamefont
  {Hildebrand}, \citenamefont {Jaouen}, \citenamefont {Kim}, \citenamefont
  {Mukherjee}, \citenamefont {Ma}, \citenamefont {Muntwiler}, \citenamefont
  {von Rohr}, \citenamefont {Cacho},\ and\ \citenamefont
  {Monney}}]{Nicholson:2021h}%
  \BibitemOpen
  \bibfield  {author} {\bibinfo {author} {\bibfnamefont {C.~W.}\ \bibnamefont
  {Nicholson}}, \bibinfo {author} {\bibfnamefont {M.}~\bibnamefont {Rumo}},
  \bibinfo {author} {\bibfnamefont {A.}~\bibnamefont {Pulkkinen}}, \bibinfo
  {author} {\bibfnamefont {G.}~\bibnamefont {Kremer}}, \bibinfo {author}
  {\bibfnamefont {B.}~\bibnamefont {Salzmann}}, \bibinfo {author}
  {\bibfnamefont {M.-L.}\ \bibnamefont {Mottas}}, \bibinfo {author}
  {\bibfnamefont {B.}~\bibnamefont {Hildebrand}}, \bibinfo {author}
  {\bibfnamefont {T.}~\bibnamefont {Jaouen}}, \bibinfo {author} {\bibfnamefont
  {T.~K.}\ \bibnamefont {Kim}}, \bibinfo {author} {\bibfnamefont
  {S.}~\bibnamefont {Mukherjee}}, \bibinfo {author} {\bibfnamefont
  {K.}~\bibnamefont {Ma}}, \bibinfo {author} {\bibfnamefont {M.}~\bibnamefont
  {Muntwiler}}, \bibinfo {author} {\bibfnamefont {F.~O.}\ \bibnamefont {von
  Rohr}}, \bibinfo {author} {\bibfnamefont {C.}~\bibnamefont {Cacho}}, \ and\
  \bibinfo {author} {\bibfnamefont {C.}~\bibnamefont {Monney}},\ }\href@noop {}
  {\bibfield  {journal} {\bibinfo  {journal} {Communications Materials}\
  }\textbf {\bibinfo {volume} {2}},\ \bibinfo {pages} {25} (\bibinfo {year}
  {2021})}\BibitemShut {NoStop}%
\bibitem [{\citenamefont {Fei}\ \emph {et~al.}(2018)\citenamefont {Fei},
  \citenamefont {Bo}, \citenamefont {Wang}, \citenamefont {Ying}, \citenamefont
  {Li}, \citenamefont {Chen}, \citenamefont {Dai}, \citenamefont {Chen},
  \citenamefont {Sun}, \citenamefont {Zhang}, \citenamefont {Qu}, \citenamefont
  {Zhang}, \citenamefont {Wang}, \citenamefont {Wang}, \citenamefont {Cao},
  \citenamefont {Bu}, \citenamefont {Song}, \citenamefont {Wan},\ and\
  \citenamefont {Wang}}]{FeiBo2018}%
  \BibitemOpen
  \bibfield  {author} {\bibinfo {author} {\bibfnamefont {F.}~\bibnamefont
  {Fei}}, \bibinfo {author} {\bibfnamefont {X.}~\bibnamefont {Bo}}, \bibinfo
  {author} {\bibfnamefont {P.}~\bibnamefont {Wang}}, \bibinfo {author}
  {\bibfnamefont {J.}~\bibnamefont {Ying}}, \bibinfo {author} {\bibfnamefont
  {J.}~\bibnamefont {Li}}, \bibinfo {author} {\bibfnamefont {K.}~\bibnamefont
  {Chen}}, \bibinfo {author} {\bibfnamefont {Q.}~\bibnamefont {Dai}}, \bibinfo
  {author} {\bibfnamefont {B.}~\bibnamefont {Chen}}, \bibinfo {author}
  {\bibfnamefont {Z.}~\bibnamefont {Sun}}, \bibinfo {author} {\bibfnamefont
  {M.}~\bibnamefont {Zhang}}, \bibinfo {author} {\bibfnamefont
  {F.}~\bibnamefont {Qu}}, \bibinfo {author} {\bibfnamefont {Y.}~\bibnamefont
  {Zhang}}, \bibinfo {author} {\bibfnamefont {Q.}~\bibnamefont {Wang}},
  \bibinfo {author} {\bibfnamefont {X.}~\bibnamefont {Wang}}, \bibinfo {author}
  {\bibfnamefont {L.}~\bibnamefont {Cao}}, \bibinfo {author} {\bibfnamefont
  {H.}~\bibnamefont {Bu}}, \bibinfo {author} {\bibfnamefont {F.}~\bibnamefont
  {Song}}, \bibinfo {author} {\bibfnamefont {X.}~\bibnamefont {Wan}}, \ and\
  \bibinfo {author} {\bibfnamefont {B.}~\bibnamefont {Wang}},\ }\href@noop {}
  {\bibfield  {journal} {\bibinfo  {journal} {Advanced Materials}\ }\textbf
  {\bibinfo {volume} {30}},\ \bibinfo {pages} {1801556} (\bibinfo {year}
  {2018})}\BibitemShut {NoStop}%
\bibitem [{\citenamefont {Jiang}\ \emph {et~al.}(2020)\citenamefont {Jiang},
  \citenamefont {Lee}, \citenamefont {Fei}, \citenamefont {Song}, \citenamefont
  {Vescovo}, \citenamefont {Kaznatcheev}, \citenamefont {Walker},\ and\
  \citenamefont {Ahn}}]{JiangLee2020}%
  \BibitemOpen
  \bibfield  {author} {\bibinfo {author} {\bibfnamefont {J.}~\bibnamefont
  {Jiang}}, \bibinfo {author} {\bibfnamefont {S.}~\bibnamefont {Lee}}, \bibinfo
  {author} {\bibfnamefont {F.}~\bibnamefont {Fei}}, \bibinfo {author}
  {\bibfnamefont {F.}~\bibnamefont {Song}}, \bibinfo {author} {\bibfnamefont
  {E.}~\bibnamefont {Vescovo}}, \bibinfo {author} {\bibfnamefont
  {K.}~\bibnamefont {Kaznatcheev}}, \bibinfo {author} {\bibfnamefont {F.~J.}\
  \bibnamefont {Walker}}, \ and\ \bibinfo {author} {\bibfnamefont {C.~H.}\
  \bibnamefont {Ahn}},\ }\href {\doibase 10.1063/5.0011549} {\bibfield
  {journal} {\bibinfo  {journal} {APL Materials}\ }\textbf {\bibinfo {volume}
  {8}},\ \bibinfo {pages} {061106} (\bibinfo {year} {2020})},\ \Eprint
  {http://arxiv.org/abs/https://doi.org/10.1063/5.0011549}
  {https://doi.org/10.1063/5.0011549} \BibitemShut {NoStop}%
\bibitem [{\citenamefont {Kiswandhi}\ \emph {et~al.}(2013)\citenamefont
  {Kiswandhi}, \citenamefont {Brooks}, \citenamefont {Cao}, \citenamefont
  {Yan}, \citenamefont {Mandrus}, \citenamefont {Jiang},\ and\ \citenamefont
  {Zhou}}]{Kiswandhi2013}%
  \BibitemOpen
  \bibfield  {author} {\bibinfo {author} {\bibfnamefont {A.}~\bibnamefont
  {Kiswandhi}}, \bibinfo {author} {\bibfnamefont {J.~S.}\ \bibnamefont
  {Brooks}}, \bibinfo {author} {\bibfnamefont {H.~B.}\ \bibnamefont {Cao}},
  \bibinfo {author} {\bibfnamefont {J.~Q.}\ \bibnamefont {Yan}}, \bibinfo
  {author} {\bibfnamefont {D.}~\bibnamefont {Mandrus}}, \bibinfo {author}
  {\bibfnamefont {Z.}~\bibnamefont {Jiang}}, \ and\ \bibinfo {author}
  {\bibfnamefont {H.~D.}\ \bibnamefont {Zhou}},\ }\href {\doibase
  10.1103/PhysRevB.87.121107} {\bibfield  {journal} {\bibinfo  {journal} {Phys.
  Rev. B}\ }\textbf {\bibinfo {volume} {87}},\ \bibinfo {pages} {121107}
  (\bibinfo {year} {2013})}\BibitemShut {NoStop}%
\bibitem [{\citenamefont {Kudo}\ \emph {et~al.}(2013)\citenamefont {Kudo},
  \citenamefont {Kobayashi}, \citenamefont {Pyon},\ and\ \citenamefont
  {Nohara}}]{KudoKobayashi2013}%
  \BibitemOpen
  \bibfield  {author} {\bibinfo {author} {\bibfnamefont {K.}~\bibnamefont
  {Kudo}}, \bibinfo {author} {\bibfnamefont {M.}~\bibnamefont {Kobayashi}},
  \bibinfo {author} {\bibfnamefont {S.}~\bibnamefont {Pyon}}, \ and\ \bibinfo
  {author} {\bibfnamefont {M.}~\bibnamefont {Nohara}},\ }\href {\doibase
  10.7566/JPSJ.82.085001} {\bibfield  {journal} {\bibinfo  {journal} {Journal
  of the Physical Society of Japan}\ }\textbf {\bibinfo {volume} {82}},\
  \bibinfo {pages} {085001} (\bibinfo {year} {2013})},\ \Eprint
  {http://arxiv.org/abs/https://doi.org/10.7566/JPSJ.82.085001}
  {https://doi.org/10.7566/JPSJ.82.085001} \BibitemShut {NoStop}%
\bibitem [{\citenamefont {Kamitani}\ \emph {et~al.}(2013)\citenamefont
  {Kamitani}, \citenamefont {Bahramy}, \citenamefont {Arita}, \citenamefont
  {Seki}, \citenamefont {Arima}, \citenamefont {Tokura},\ and\ \citenamefont
  {Ishiwata}}]{Kamitani2013}%
  \BibitemOpen
  \bibfield  {author} {\bibinfo {author} {\bibfnamefont {M.}~\bibnamefont
  {Kamitani}}, \bibinfo {author} {\bibfnamefont {M.~S.}\ \bibnamefont
  {Bahramy}}, \bibinfo {author} {\bibfnamefont {R.}~\bibnamefont {Arita}},
  \bibinfo {author} {\bibfnamefont {S.}~\bibnamefont {Seki}}, \bibinfo {author}
  {\bibfnamefont {T.}~\bibnamefont {Arima}}, \bibinfo {author} {\bibfnamefont
  {Y.}~\bibnamefont {Tokura}}, \ and\ \bibinfo {author} {\bibfnamefont
  {S.}~\bibnamefont {Ishiwata}},\ }\href {\doibase 10.1103/PhysRevB.87.180501}
  {\bibfield  {journal} {\bibinfo  {journal} {Phys. Rev. B}\ }\textbf {\bibinfo
  {volume} {87}},\ \bibinfo {pages} {180501} (\bibinfo {year}
  {2013})}\BibitemShut {NoStop}%
\bibitem [{\citenamefont {Oh}\ \emph {et~al.}(2013)\citenamefont {Oh},
  \citenamefont {Yang}, \citenamefont {Horibe},\ and\ \citenamefont
  {Cheong}}]{OhYangHoribe2013}%
  \BibitemOpen
  \bibfield  {author} {\bibinfo {author} {\bibfnamefont {Y.~S.}\ \bibnamefont
  {Oh}}, \bibinfo {author} {\bibfnamefont {J.~J.}\ \bibnamefont {Yang}},
  \bibinfo {author} {\bibfnamefont {Y.}~\bibnamefont {Horibe}}, \ and\ \bibinfo
  {author} {\bibfnamefont {S.-W.}\ \bibnamefont {Cheong}},\ }\href {\doibase
  10.1103/PhysRevLett.110.127209} {\bibfield  {journal} {\bibinfo  {journal}
  {Phys. Rev. Lett.}\ }\textbf {\bibinfo {volume} {110}},\ \bibinfo {pages}
  {127209} (\bibinfo {year} {2013})}\BibitemShut {NoStop}%
\bibitem [{\citenamefont {Pascut}\ \emph
  {et~al.}(2014{\natexlab{a}})\citenamefont {Pascut}, \citenamefont {Birol},
  \citenamefont {Gutmann}, \citenamefont {Yang}, \citenamefont {Cheong},
  \citenamefont {Haule},\ and\ \citenamefont {Kiryukhin}}]{PascutBirol2014}%
  \BibitemOpen
  \bibfield  {author} {\bibinfo {author} {\bibfnamefont {G.~L.}\ \bibnamefont
  {Pascut}}, \bibinfo {author} {\bibfnamefont {T.}~\bibnamefont {Birol}},
  \bibinfo {author} {\bibfnamefont {M.~J.}\ \bibnamefont {Gutmann}}, \bibinfo
  {author} {\bibfnamefont {J.~J.}\ \bibnamefont {Yang}}, \bibinfo {author}
  {\bibfnamefont {S.-W.}\ \bibnamefont {Cheong}}, \bibinfo {author}
  {\bibfnamefont {K.}~\bibnamefont {Haule}}, \ and\ \bibinfo {author}
  {\bibfnamefont {V.}~\bibnamefont {Kiryukhin}},\ }\href {\doibase
  10.1103/PhysRevB.90.195122} {\bibfield  {journal} {\bibinfo  {journal} {Phys.
  Rev. B}\ }\textbf {\bibinfo {volume} {90}},\ \bibinfo {pages} {195122}
  (\bibinfo {year} {2014}{\natexlab{a}})}\BibitemShut {NoStop}%
\bibitem [{\citenamefont {Saleh}\ and\ \citenamefont
  {Artyukhin}(2020)}]{Saleh2020}%
  \BibitemOpen
  \bibfield  {author} {\bibinfo {author} {\bibfnamefont {G.}~\bibnamefont
  {Saleh}}\ and\ \bibinfo {author} {\bibfnamefont {S.}~\bibnamefont
  {Artyukhin}},\ }\href {\doibase 10.1021/acs.jpclett.0c00012} {\bibfield
  {journal} {\bibinfo  {journal} {The Journal of Physical Chemistry Letters}\
  }\textbf {\bibinfo {volume} {11}},\ \bibinfo {pages} {2127} (\bibinfo {year}
  {2020})},\ \bibinfo {note} {pMID: 32079398},\ \Eprint
  {http://arxiv.org/abs/https://doi.org/10.1021/acs.jpclett.0c00012}
  {https://doi.org/10.1021/acs.jpclett.0c00012} \BibitemShut {NoStop}%
\bibitem [{\citenamefont {Hsu}\ \emph {et~al.}(2013)\citenamefont {Hsu},
  \citenamefont {Mauerer}, \citenamefont {Vogt}, \citenamefont {Yang},
  \citenamefont {Oh}, \citenamefont {Cheong}, \citenamefont {Bode},\ and\
  \citenamefont {Wu}}]{Hsu2013}%
  \BibitemOpen
  \bibfield  {author} {\bibinfo {author} {\bibfnamefont {P.-J.}\ \bibnamefont
  {Hsu}}, \bibinfo {author} {\bibfnamefont {T.}~\bibnamefont {Mauerer}},
  \bibinfo {author} {\bibfnamefont {M.}~\bibnamefont {Vogt}}, \bibinfo {author}
  {\bibfnamefont {J.~J.}\ \bibnamefont {Yang}}, \bibinfo {author}
  {\bibfnamefont {Y.~S.}\ \bibnamefont {Oh}}, \bibinfo {author} {\bibfnamefont
  {S.-W.}\ \bibnamefont {Cheong}}, \bibinfo {author} {\bibfnamefont
  {M.}~\bibnamefont {Bode}}, \ and\ \bibinfo {author} {\bibfnamefont
  {W.}~\bibnamefont {Wu}},\ }\href {\doibase 10.1103/PhysRevLett.111.266401}
  {\bibfield  {journal} {\bibinfo  {journal} {Phys. Rev. Lett.}\ }\textbf
  {\bibinfo {volume} {111}},\ \bibinfo {pages} {266401} (\bibinfo {year}
  {2013})}\BibitemShut {NoStop}%
\bibitem [{\citenamefont {Chen}\ \emph {et~al.}(2017)\citenamefont {Chen},
  \citenamefont {Kim}, \citenamefont {Yang}, \citenamefont {Cao}, \citenamefont
  {Jin},\ and\ \citenamefont {Plummer}}]{Chen:2017i}%
  \BibitemOpen
  \bibfield  {author} {\bibinfo {author} {\bibfnamefont {C.}~\bibnamefont
  {Chen}}, \bibinfo {author} {\bibfnamefont {J.}~\bibnamefont {Kim}}, \bibinfo
  {author} {\bibfnamefont {Y.}~\bibnamefont {Yang}}, \bibinfo {author}
  {\bibfnamefont {G.}~\bibnamefont {Cao}}, \bibinfo {author} {\bibfnamefont
  {R.}~\bibnamefont {Jin}}, \ and\ \bibinfo {author} {\bibfnamefont {E.~W.}\
  \bibnamefont {Plummer}},\ }\href@noop {} {\bibfield  {journal} {\bibinfo
  {journal} {Phys. Rev. B}\ }\textbf {\bibinfo {volume} {95}},\ \bibinfo
  {pages} {094118} (\bibinfo {year} {2017})}\BibitemShut {NoStop}%
\bibitem [{\citenamefont {Park}\ \emph {et~al.}(2021)\citenamefont {Park},
  \citenamefont {Kim}, \citenamefont {Kim}, \citenamefont {Kim}, \citenamefont
  {Kim}, \citenamefont {Kim}, \citenamefont {Choi}, \citenamefont {Won},
  \citenamefont {Kim}, \citenamefont {Kim}, \citenamefont {Talantsev},
  \citenamefont {Watanabe}, \citenamefont {Taniguchi}, \citenamefont {Cheong},
  \citenamefont {Kim}, \citenamefont {Yeom}, \citenamefont {Kim}, \citenamefont
  {Kim},\ and\ \citenamefont {Kim}}]{Park:2021f}%
  \BibitemOpen
  \bibfield  {author} {\bibinfo {author} {\bibfnamefont {S.}~\bibnamefont
  {Park}}, \bibinfo {author} {\bibfnamefont {S.~Y.}\ \bibnamefont {Kim}},
  \bibinfo {author} {\bibfnamefont {H.~K.}\ \bibnamefont {Kim}}, \bibinfo
  {author} {\bibfnamefont {M.~J.}\ \bibnamefont {Kim}}, \bibinfo {author}
  {\bibfnamefont {T.}~\bibnamefont {Kim}}, \bibinfo {author} {\bibfnamefont
  {H.}~\bibnamefont {Kim}}, \bibinfo {author} {\bibfnamefont {G.~S.}\
  \bibnamefont {Choi}}, \bibinfo {author} {\bibfnamefont {C.~J.}\ \bibnamefont
  {Won}}, \bibinfo {author} {\bibfnamefont {S.}~\bibnamefont {Kim}}, \bibinfo
  {author} {\bibfnamefont {K.}~\bibnamefont {Kim}}, \bibinfo {author}
  {\bibfnamefont {E.~F.}\ \bibnamefont {Talantsev}}, \bibinfo {author}
  {\bibfnamefont {K.}~\bibnamefont {Watanabe}}, \bibinfo {author}
  {\bibfnamefont {T.}~\bibnamefont {Taniguchi}}, \bibinfo {author}
  {\bibfnamefont {S.-W.}\ \bibnamefont {Cheong}}, \bibinfo {author}
  {\bibfnamefont {B.~J.}\ \bibnamefont {Kim}}, \bibinfo {author} {\bibfnamefont
  {H.~W.}\ \bibnamefont {Yeom}}, \bibinfo {author} {\bibfnamefont
  {J.}~\bibnamefont {Kim}}, \bibinfo {author} {\bibfnamefont {T.-H.}\
  \bibnamefont {Kim}}, \ and\ \bibinfo {author} {\bibfnamefont {J.~S.}\
  \bibnamefont {Kim}},\ }\href@noop {} {\bibfield  {journal} {\bibinfo
  {journal} {Nature Communications}\ }\textbf {\bibinfo {volume} {12}},\
  \bibinfo {pages} {3157} (\bibinfo {year} {2021})}\BibitemShut {NoStop}%
\bibitem [{\citenamefont {Yang}\ \emph {et~al.}(2012)\citenamefont {Yang},
  \citenamefont {Choi}, \citenamefont {Oh}, \citenamefont {Hogan},
  \citenamefont {Horibe}, \citenamefont {Kim}, \citenamefont {Min},\ and\
  \citenamefont {Cheong}}]{Yang:2012a}%
  \BibitemOpen
  \bibfield  {author} {\bibinfo {author} {\bibfnamefont {J.}~\bibnamefont
  {Yang}}, \bibinfo {author} {\bibfnamefont {Y.}~\bibnamefont {Choi}}, \bibinfo
  {author} {\bibfnamefont {Y.}~\bibnamefont {Oh}}, \bibinfo {author}
  {\bibfnamefont {A.}~\bibnamefont {Hogan}}, \bibinfo {author} {\bibfnamefont
  {Y.}~\bibnamefont {Horibe}}, \bibinfo {author} {\bibfnamefont
  {K.}~\bibnamefont {Kim}}, \bibinfo {author} {\bibfnamefont {B.}~\bibnamefont
  {Min}}, \ and\ \bibinfo {author} {\bibfnamefont {S.}~\bibnamefont {Cheong}},\
  }\href@noop {} {\bibfield  {journal} {\bibinfo  {journal} {Physical Review
  Letters}\ }\textbf {\bibinfo {volume} {108}},\ \bibinfo {pages} {116402}
  (\bibinfo {year} {2012})}\BibitemShut {NoStop}%
\bibitem [{\citenamefont {Ootsuki}\ \emph {et~al.}(2012)\citenamefont
  {Ootsuki}, \citenamefont {Wakisaka}, \citenamefont {Pyon}, \citenamefont
  {Kudo}, \citenamefont {Nohara}, \citenamefont {Arita}, \citenamefont {Anzai},
  \citenamefont {Namatame}, \citenamefont {Taniguchi}, \citenamefont {Saini},\
  and\ \citenamefont {Mizokawa}}]{Ootsuki2012}%
  \BibitemOpen
  \bibfield  {author} {\bibinfo {author} {\bibfnamefont {D.}~\bibnamefont
  {Ootsuki}}, \bibinfo {author} {\bibfnamefont {Y.}~\bibnamefont {Wakisaka}},
  \bibinfo {author} {\bibfnamefont {S.}~\bibnamefont {Pyon}}, \bibinfo {author}
  {\bibfnamefont {K.}~\bibnamefont {Kudo}}, \bibinfo {author} {\bibfnamefont
  {M.}~\bibnamefont {Nohara}}, \bibinfo {author} {\bibfnamefont
  {M.}~\bibnamefont {Arita}}, \bibinfo {author} {\bibfnamefont
  {H.}~\bibnamefont {Anzai}}, \bibinfo {author} {\bibfnamefont
  {H.}~\bibnamefont {Namatame}}, \bibinfo {author} {\bibfnamefont
  {M.}~\bibnamefont {Taniguchi}}, \bibinfo {author} {\bibfnamefont {N.~L.}\
  \bibnamefont {Saini}}, \ and\ \bibinfo {author} {\bibfnamefont
  {T.}~\bibnamefont {Mizokawa}},\ }\href {\doibase 10.1103/PhysRevB.86.014519}
  {\bibfield  {journal} {\bibinfo  {journal} {Phys. Rev. B}\ }\textbf {\bibinfo
  {volume} {86}},\ \bibinfo {pages} {014519} (\bibinfo {year}
  {2012})}\BibitemShut {NoStop}%
\bibitem [{\citenamefont {Qian}\ \emph {et~al.}(2014)\citenamefont {Qian},
  \citenamefont {Miao}, \citenamefont {Wang}, \citenamefont {Shi},
  \citenamefont {Huang}, \citenamefont {Zhang}, \citenamefont {Xu},
  \citenamefont {Zeng}, \citenamefont {Ma}, \citenamefont {Richard},
  \citenamefont {Shi}, \citenamefont {Xu}, \citenamefont {Dai}, \citenamefont
  {Fang}, \citenamefont {Fang}, \citenamefont {Wang},\ and\ \citenamefont
  {Ding}}]{Qian2014}%
  \BibitemOpen
  \bibfield  {author} {\bibinfo {author} {\bibfnamefont {T.}~\bibnamefont
  {Qian}}, \bibinfo {author} {\bibfnamefont {H.}~\bibnamefont {Miao}}, \bibinfo
  {author} {\bibfnamefont {Z.~J.}\ \bibnamefont {Wang}}, \bibinfo {author}
  {\bibfnamefont {X.}~\bibnamefont {Shi}}, \bibinfo {author} {\bibfnamefont
  {Y.~B.}\ \bibnamefont {Huang}}, \bibinfo {author} {\bibfnamefont
  {P.}~\bibnamefont {Zhang}}, \bibinfo {author} {\bibfnamefont
  {N.}~\bibnamefont {Xu}}, \bibinfo {author} {\bibfnamefont {L.~K.}\
  \bibnamefont {Zeng}}, \bibinfo {author} {\bibfnamefont {J.~Z.}\ \bibnamefont
  {Ma}}, \bibinfo {author} {\bibfnamefont {P.}~\bibnamefont {Richard}},
  \bibinfo {author} {\bibfnamefont {M.}~\bibnamefont {Shi}}, \bibinfo {author}
  {\bibfnamefont {G.}~\bibnamefont {Xu}}, \bibinfo {author} {\bibfnamefont
  {X.}~\bibnamefont {Dai}}, \bibinfo {author} {\bibfnamefont {Z.}~\bibnamefont
  {Fang}}, \bibinfo {author} {\bibfnamefont {A.~F.}\ \bibnamefont {Fang}},
  \bibinfo {author} {\bibfnamefont {N.~L.}\ \bibnamefont {Wang}}, \ and\
  \bibinfo {author} {\bibfnamefont {H.}~\bibnamefont {Ding}},\ }\href
  {http://stacks.iop.org/1367-2630/16/i=12/a=123038} {\bibfield  {journal}
  {\bibinfo  {journal} {New Journal of Physics}\ }\textbf {\bibinfo {volume}
  {16}},\ \bibinfo {pages} {123038} (\bibinfo {year} {2014})}\BibitemShut
  {NoStop}%
\bibitem [{\citenamefont {Pascut}\ \emph
  {et~al.}(2014{\natexlab{b}})\citenamefont {Pascut}, \citenamefont {Haule},
  \citenamefont {Gutmann}, \citenamefont {Barnett}, \citenamefont {Bombardi},
  \citenamefont {Artyukhin}, \citenamefont {Birol}, \citenamefont {Vanderbilt},
  \citenamefont {Yang}, \citenamefont {Cheong},\ and\ \citenamefont
  {Kiryukhin}}]{PascutHaule2014}%
  \BibitemOpen
  \bibfield  {author} {\bibinfo {author} {\bibfnamefont {G.~L.}\ \bibnamefont
  {Pascut}}, \bibinfo {author} {\bibfnamefont {K.}~\bibnamefont {Haule}},
  \bibinfo {author} {\bibfnamefont {M.~J.}\ \bibnamefont {Gutmann}}, \bibinfo
  {author} {\bibfnamefont {S.~A.}\ \bibnamefont {Barnett}}, \bibinfo {author}
  {\bibfnamefont {A.}~\bibnamefont {Bombardi}}, \bibinfo {author}
  {\bibfnamefont {S.}~\bibnamefont {Artyukhin}}, \bibinfo {author}
  {\bibfnamefont {T.}~\bibnamefont {Birol}}, \bibinfo {author} {\bibfnamefont
  {D.}~\bibnamefont {Vanderbilt}}, \bibinfo {author} {\bibfnamefont {J.~J.}\
  \bibnamefont {Yang}}, \bibinfo {author} {\bibfnamefont {S.-W.}\ \bibnamefont
  {Cheong}}, \ and\ \bibinfo {author} {\bibfnamefont {V.}~\bibnamefont
  {Kiryukhin}},\ }\href {\doibase 10.1103/PhysRevLett.112.086402} {\bibfield
  {journal} {\bibinfo  {journal} {Phys. Rev. Lett.}\ }\textbf {\bibinfo
  {volume} {112}},\ \bibinfo {pages} {086402} (\bibinfo {year}
  {2014}{\natexlab{b}})}\BibitemShut {NoStop}%
\bibitem [{\citenamefont {Ku}\ \emph {et~al.}(2010)\citenamefont {Ku},
  \citenamefont {Berlijn},\ and\ \citenamefont {Lee}}]{KuPRL2010}%
  \BibitemOpen
  \bibfield  {author} {\bibinfo {author} {\bibfnamefont {W.}~\bibnamefont
  {Ku}}, \bibinfo {author} {\bibfnamefont {T.}~\bibnamefont {Berlijn}}, \ and\
  \bibinfo {author} {\bibfnamefont {C.-C.}\ \bibnamefont {Lee}},\ }\href
  {\doibase 10.1103/PhysRevLett.104.216401} {\bibfield  {journal} {\bibinfo
  {journal} {Phys. Rev. Lett.}\ }\textbf {\bibinfo {volume} {104}},\ \bibinfo
  {pages} {216401} (\bibinfo {year} {2010})}\BibitemShut {NoStop}%
\bibitem [{\citenamefont {{Rigaku Oxford Diffraction}}(2018)}]{CrysAlis}%
  \BibitemOpen
  \bibfield  {author} {\bibinfo {author} {\bibnamefont {{Rigaku Oxford
  Diffraction}}},\ }\href@noop {} {\enquote {\bibinfo {title} {Crys{A}lis{P}ro
  {S}oftware system version 1.171.39.46, {R}igaku {C}orporation, {O}xford},}\ }
  (\bibinfo {year} {2018})\BibitemShut {NoStop}%
\bibitem [{\citenamefont {Sheldrick}(2015{\natexlab{a}})}]{Sheldrick2015a}%
  \BibitemOpen
  \bibfield  {author} {\bibinfo {author} {\bibfnamefont {G.~M.}\ \bibnamefont
  {Sheldrick}},\ }\href {\doibase 10.1107/s2053273314026370} {\bibfield
  {journal} {\bibinfo  {journal} {Acta Crystallographica Section A Foundations
  and Advances}\ }\textbf {\bibinfo {volume} {71}},\ \bibinfo {pages} {3}
  (\bibinfo {year} {2015}{\natexlab{a}})}\BibitemShut {NoStop}%
\bibitem [{\citenamefont {Sheldrick}(2015{\natexlab{b}})}]{Sheldrick2015}%
  \BibitemOpen
  \bibfield  {author} {\bibinfo {author} {\bibfnamefont {G.~M.}\ \bibnamefont
  {Sheldrick}},\ }\href {\doibase 10.1107/s2053229614024218} {\bibfield
  {journal} {\bibinfo  {journal} {Acta Crystallographica Section C Structural
  Chemistry}\ }\textbf {\bibinfo {volume} {71}},\ \bibinfo {pages} {3}
  (\bibinfo {year} {2015}{\natexlab{b}})}\BibitemShut {NoStop}%
\bibitem [{\citenamefont {Farrugia}(2012)}]{Farrugia2012}%
  \BibitemOpen
  \bibfield  {author} {\bibinfo {author} {\bibfnamefont {L.~J.}\ \bibnamefont
  {Farrugia}},\ }\href {\doibase 10.1107/s0021889812029111} {\bibfield
  {journal} {\bibinfo  {journal} {Journal of Applied Crystallography}\ }\textbf
  {\bibinfo {volume} {45}},\ \bibinfo {pages} {849} (\bibinfo {year}
  {2012})}\BibitemShut {NoStop}%
\bibitem [{\citenamefont {Koepernik}\ and\ \citenamefont
  {Eschrig}(1999)}]{Koepernik1999}%
  \BibitemOpen
  \bibfield  {author} {\bibinfo {author} {\bibfnamefont {K.}~\bibnamefont
  {Koepernik}}\ and\ \bibinfo {author} {\bibfnamefont {H.}~\bibnamefont
  {Eschrig}},\ }\href {\doibase 10.1103/PhysRevB.59.1743} {\bibfield  {journal}
  {\bibinfo  {journal} {Phys. Rev. B}\ }\textbf {\bibinfo {volume} {59}},\
  \bibinfo {pages} {1743} (\bibinfo {year} {1999})}\BibitemShut {NoStop}%
\bibitem [{\citenamefont {Perdew}\ and\ \citenamefont
  {Wang}(1992)}]{Perdew1992}%
  \BibitemOpen
  \bibfield  {author} {\bibinfo {author} {\bibfnamefont {J.~P.}\ \bibnamefont
  {Perdew}}\ and\ \bibinfo {author} {\bibfnamefont {Y.}~\bibnamefont {Wang}},\
  }\href {\doibase 10.1103/PhysRevB.45.13244} {\bibfield  {journal} {\bibinfo
  {journal} {Phys. Rev. B}\ }\textbf {\bibinfo {volume} {45}},\ \bibinfo
  {pages} {13244} (\bibinfo {year} {1992})}\BibitemShut {NoStop}%
\bibitem [{\citenamefont {Perdew}\ \emph {et~al.}(1996)\citenamefont {Perdew},
  \citenamefont {Burke},\ and\ \citenamefont
  {Ernzerhof}}]{PerdewBurkeErnzerhof1996}%
  \BibitemOpen
  \bibfield  {author} {\bibinfo {author} {\bibfnamefont {J.~P.}\ \bibnamefont
  {Perdew}}, \bibinfo {author} {\bibfnamefont {K.}~\bibnamefont {Burke}}, \
  and\ \bibinfo {author} {\bibfnamefont {M.}~\bibnamefont {Ernzerhof}},\ }\href
  {\doibase 10.1103/PhysRevLett.77.3865} {\bibfield  {journal} {\bibinfo
  {journal} {Phys. Rev. Lett.}\ }\textbf {\bibinfo {volume} {77}},\ \bibinfo
  {pages} {3865} (\bibinfo {year} {1996})}\BibitemShut {NoStop}%
\bibitem [{\citenamefont {Eschrig}\ \emph {et~al.}(2004)\citenamefont
  {Eschrig}, \citenamefont {Richter},\ and\ \citenamefont
  {Opahle}}]{Eschrig2004}%
  \BibitemOpen
  \bibfield  {author} {\bibinfo {author} {\bibfnamefont {H.}~\bibnamefont
  {Eschrig}}, \bibinfo {author} {\bibfnamefont {M.}~\bibnamefont {Richter}}, \
  and\ \bibinfo {author} {\bibfnamefont {I.}~\bibnamefont {Opahle}},\
  }\href@noop {} {\emph {\bibinfo {title} {Relativistic Solid State
  Calculations, in: Relativistic Electronic Structure Theory, Part 2.
  Applications}}},\ edited by\ \bibinfo {editor} {\bibfnamefont
  {P.}~\bibnamefont {Schwerdtfeger}},\ Vol.~\bibinfo {volume} {13}\ (\bibinfo
  {publisher} {Elsevier},\ \bibinfo {year} {2004})\ pp.\ \bibinfo {pages}
  {723--776}\BibitemShut {NoStop}%
\end{thebibliography}
\end{document}